\documentclass[a5paper, 10pt]{article}

\usepackage{cite, amsmath, amssymb, booktabs, lastpage, graphicx}
\usepackage[hidelinks]{hyperref}

\usepackage{geometry}
\usepackage{setspace}

\usepackage{amsthm}

\theoremstyle{plain}

\theoremstyle{remark}

\theoremstyle{definition}

\usepackage{graphicx}
\makeatletter
\renewcommand{\maketitle}{
	\begin{center}
    \rule[.2em]{\textwidth}{0.353mm}
		\begin{minipage}[m]{0.35\textwidth}
			{\scriptsize
				\begin{center}
					\textsf{\textbf{\huge MATCH}\\
						\textit{Communications in Mathematical\\
							and in Computer Chemistry}
					}
			\end{center}}
		\end{minipage}\hfill
		\begin{minipage}[m]{0.65\textwidth}
			\begin{flushright}
				\baselineskip=10px
				{\scriptsize\sffamily{\itshape  MATCH Commun. Math. Comput. Chem.}
				\vol{}
				(\pubyear)
					}\\
				{\scriptsize\sffamily {\bfseries ISSN:} 0340--6253}\\
				{\scriptsize\sffamily \textbf{doi:} \doi{10.46793/match.94-1.097A.}}
			\end{flushright}
		\end{minipage}
		\rule[1em]{\textwidth}{.353mm}
		\baselineskip=0.30in
		{\Large\bfseries \@title} \par
		\vspace{5mm}
		\baselineskip=0.2in
		{\large\bfseries \@author}\par
		\vspace{1mm}
		{\it \@address} \par
		{\small\tt \@email} \par
		\vspace{3mm}
		{\small (Accepted December 20, 2024 )} \par
	\end{center}
	\vspace{3mm}
}
\newcommand{\address}[1]{\def\@address{#1}}
\newcommand{\email}[1]{\def\@email{#1}}
\newcommand{\doi}[1]{\href{https://doi.org/#1}{#1}}
\date{November 12, 2024}

\newcommand{\vol}{\textbf{94}}
\newcommand{\pubyear}{2025}

\makeatother

\newcommand{\acknowledgment}[1]{\vspace{5mm}\singlespacing
	{\noindent\textbf{\textit{Acknowledgment\/}:} #1}
}

\usepackage{fancyhdr}

\usepackage[margin=1cm,%
			font=footnotesize,%
			format=hang,%
			labelsep=period,%
			labelfont=bf]{caption}
    
\usepackage{amsthm,graphicx,mathtools,url,hyperref,lineno,verbatim,soul,tabulary,authblk}
\newcolumntype{C}[1]{>{\centering\arraybackslash}p{#1}}
\usepackage{physics}
\newtheorem{dfn}{Definition}[section]
\newtheorem{pro}[dfn]{Problem}
\newtheorem{thm}[dfn]{Theorem}
\newtheorem{exa}[dfn]{Example}

\newtheorem{prot}[dfn]{Protocol}
\newtheorem{cor}[dfn]{Corollary}
\newtheorem{lem}[dfn]{Lemma}
\newtheorem{prop}[dfn]{Proposition}

\newcommand{\C}{\mathbb C}
\newcommand{\R}{\mathbb R}

\newcommand{\al}{\alpha}
\newcommand{\be}{\beta}

\newcommand{\de}{\delta}

\newcommand{\la}{\lambda}

\newcommand{\Si}{\Sigma}

\newcommand{\ep}{\varepsilon}
\newcommand{\ph}{\varphi}
\newcommand{\angstrom}{\textup{\AA}}
\newcommand{\RMSD}{\mathrm{RMSD}}

\newcommand{\SO}{\mathrm{SO}}
\newcommand{\Or}{\mathrm{O}}

\newcommand{\trin}{\mathrm{TRIN}}
\newcommand{\bri}{\mathrm{BRI}}
\newcommand{\brain}{\mathrm{Brain}}

\newcommand{\bid}{\mathrm{BID}}
\newcommand{\bib}{\mathrm{BIB}}
\newcommand{\bris}{\mathrm{BRIS}}

\newcommand{\TM}{\mathrm{TM}}
\newcommand{\TMD}{\mathrm{TMD}}

\newcommand{\ve}[1]{\overrightarrow{#1}}
\newcommand{\ov}[1]{\overline{#1}}

\graphicspath{{images/}}

\title{A Complete and Bi-Continuous Invariant of Protein Backbones under Rigid Motion}
\author{Olga Anosova$^a$, Alexey Gorelov$^b$, William Jeffcott$^a$, 
Ziqiu Jiang$^c$, Vitaliy Kurlin$^{a,}$\footnote{Corresponding author.}}

\address{$^a$Computer Science, University of Liverpool, Liverpool, L69 3BX, UK\\
	$^b$Université Grenoble Alpes, Institut Fourier, 38000 Grenoble, France\\
	$^c$Department of Surgery \& Cancer, Faculty of Medicine, \\
Imperial College London, London, W12 0NN, UK
    }

\email{vitaliy.kurlin@liverpool.ac.uk}

\date{\today}

\begin{document}

\maketitle

\begin{abstract}
Proteins are large biomolecules that regulate all living organisms and 
consist of one or several chains.
The \emph{primary} structure of a protein chain is a sequence of amino acid residues whose three main atoms (alpha-carbon, nitrogen, and carbonyl carbon) form a protein backbone.
The \emph{tertiary} structure is the rigid shape of a protein chain represented by atomic positions in 3-dimensional space.
Since different geometric structures often have distinct functional properties, it is important to continuously quantify differences in rigid shapes of protein backbones.
Unfortunately, many widely used similarities of proteins fail axioms of a distance metric and discontinuously change under tiny perturbations of atoms. 
\medskip

This paper develops a complete invariant that identifies any protein backbone in 3-dimensional space, uniquely under rigid motion. 
This invariant is Lipschitz bi-continuous in the sense that it changes up to a constant multiple of a maximum perturbation of atoms, and vice versa. 
The new invariant has been used to detect thousands of (near-)duplicates in the Protein Data Bank, whose presence inevitably skews machine learning predictions.  
The resulting invariant space allows low-dimensional maps with analytically defined coordinates that reveal substantial variability in the protein universe.
\end{abstract}


\section{Motivations and the problem statement}
\label{sec:intro}

\begin{figure}[h!]
\includegraphics[height=21mm]{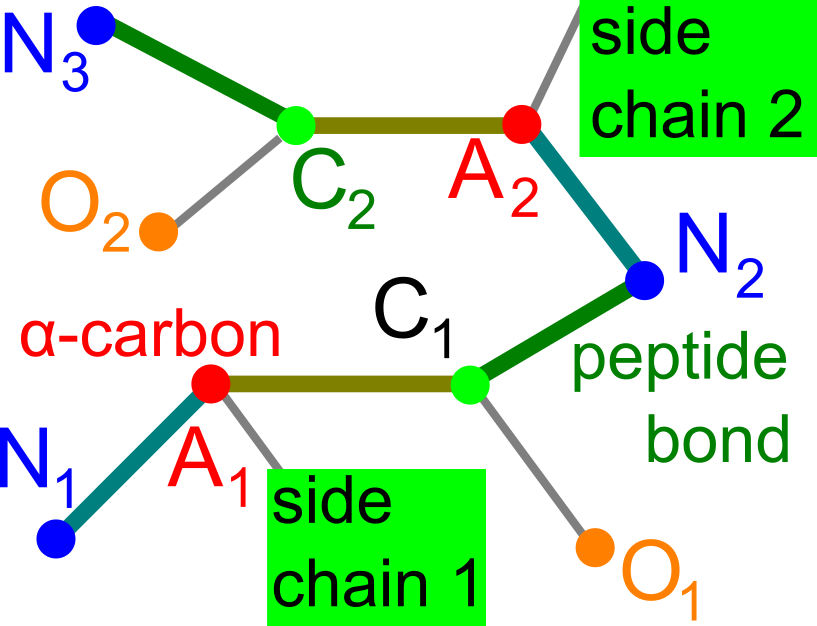}
\hspace*{1mm}
\includegraphics[height=21mm]{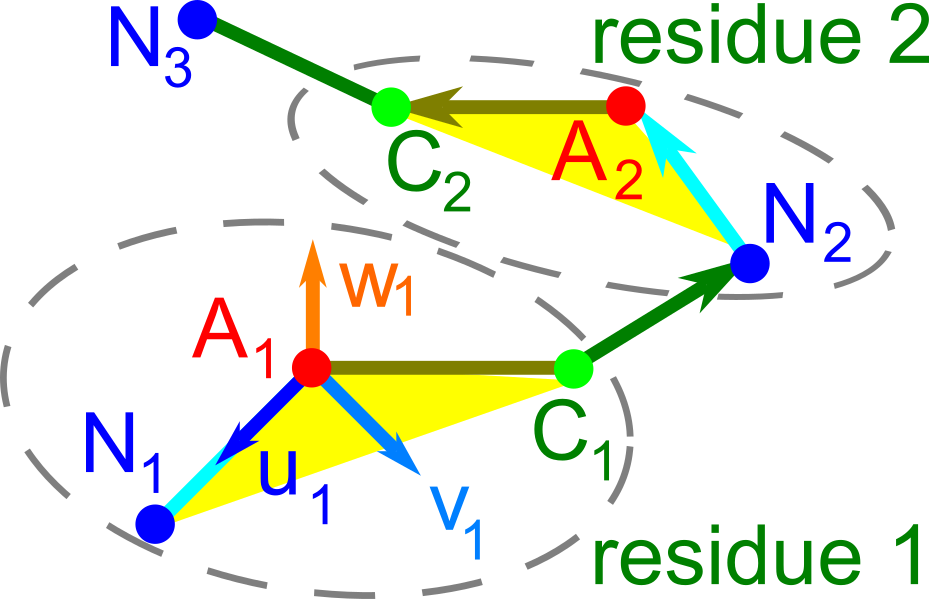}
\hspace*{1mm}
\includegraphics[height=21mm]{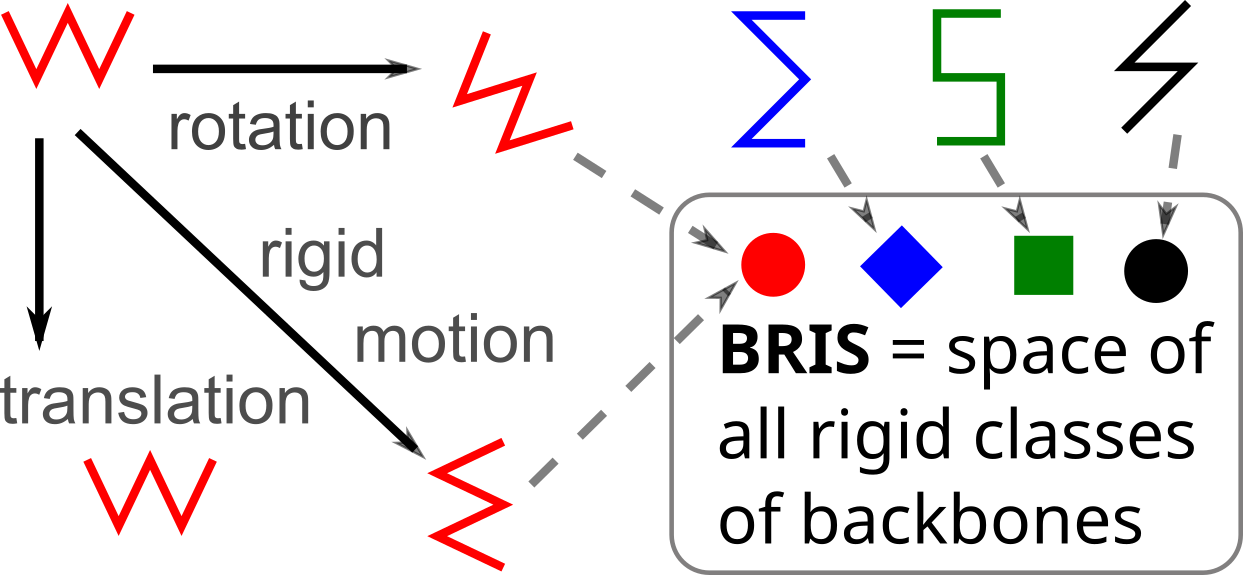}
\caption{\textbf{Left}: 
all main atoms $N_i$, $A_i$, $C_i$  of a protein chain form a \emph{backbone} embedded in $\R^3$.
\textbf{Middle}: each triangle $\triangle N_i A_i C_i$ defines an orthonormal basis $\vb*{u}_i,\vb*{v}_i,\vb*{w}_i$.
The coordinates of the bonds $\ve{C_i N_{i+1}}$, $\ve{N_{i+1}A_{i+1}}$,  $\ve{A_{i+1}C_{i+1}}$ in this basis form the complete Backbone Rigid Invariant $\bri$.
\textbf{Right}: 
All rigidly equivalent backbones form a single \emph{rigid class}.
All rigid classes of backbones form the \emph{Backbone Rigid Space}. 
}
\label{fig:backbones}
\end{figure}

A \emph{protein} is a large biomolecule consisting of one or several chains of amino acid residues.
The \emph{primary structure} (\emph{sequence}) of a protein chain is a string of residue labels (represented by one or three letters), each denoting one of (usually) 20 standard amino acids \cite{needleman2012protein}.
The \emph{secondary} structure consists of frequent semi-rigid subchains such as $\al$-helices and $\be$-strands \cite{linderstrom1952lane}. 
A sequence of a protein is easy to experimentally determine but important functional properties  such as interactions with drug molecules depend on a 3-dimensional geometric shape (a \emph{tertiary structure} or \emph{fold}) represented by an embedding of all its atoms in $\R^3$ \cite{scott2017mathematical}, 
see Fig.~\ref{fig:backbones}~(left). 
\medskip

In 1973, Nobel laureate Anfinsen conjectured that the sequence of any protein chain determines its 3D geometric shape \cite{anfinsen1973principles}.
Following this conjecture, neural networks such as AlphaFold2 and RosettaFold \cite{jumper2021highly,baek2021accurate,mirdita2022colabfold,van2024fast} optimize millions of parameters to predict a protein fold from its sequence but need re-training \cite{jones2022impact} on the growing number of experimental structures in the Protein Data Bank (PDB) \cite{burley2017protein}.
The reported accuracies of prediction are based on the LDDT (Local Distance Difference Test) \cite[p.~2728]{mariani2013lddt} and TM-score \cite{zhang2004scoring}, which fail the metric axioms.
Then clustering algorithms 
can produce pre-determined clusters and may not be trustworthy \cite{rass2024metricizing}.
\medskip

Backbones of the same length (number of residues) can be optimally aligned to minimize the Root Mean Square Deviation (RMSD) between corresponding atoms \cite{holm2024dali}.
This RMSD is slow to compute for all pairs of proteins and gives only distances without mapping the protein universe. 
\medskip

We develop a different approach by mapping the space of protein backbones in analytically defined coordinates similar to geographic-style maps of a new planet.  
The first question that we should ask about any real data such as protein tertiary structures is ``same or different'' \cite{sacchi2020same}.
\medskip

Any embedded protein in $\R^3$ can be rigidly moved (translated or rotated), which changes all atomic coordinates but the underlying structure remains the same in the sense that different images of a protein under rigid motion have the same properties in a fixed environment. 
Though proteins are flexible molecules, it is important to distinguish their rigid shapes that can differently interact \cite{heifetz2003effect} with other molecules including medical drugs.

\begin{dfn}[Backbone Rigid Space $\bris_m$]
\label{dfn:bris}
A protein \emph{backbone} is a sequence of $m$ ordered triplets of main chain atoms (nitrogen $N_i$, $\al$-carbon $A_i$, and carbonyl carbon $C_i$) given by their geometric positions in $\R^3$.
A rigid \emph{motion} is a composition of translations and rotations matching backbones in $\R^3$ (denoted by $S\cong Q$). 
The classes of all backbones of $m$ triplets under rigid motion form the \emph{Backbone Rigid Space} $\bris_m$.
\end{dfn}

Rigid classes of backbones can be distinguished only by an \emph{invariant} $I$ defined as a descriptor preserved under any rigid motion.
Any non-invariant descriptor $J$ always has a \emph{false negative} pair of backbones $S\cong Q$ with $J(S)\neq J(Q)$.
The number of residues is invariant, while the center of mass moves together with a backbone and is not invariant. 
\medskip

Backbones were studied by incomplete invariants such as torsion angles, which allow \emph{false positive} pairs of non-equivalent backbones $S\not\cong Q$ with $I(S)=I(Q)$.
Since all atoms in a backbone $S$ are ordered, their distance matrix determines $S\subset\R^3$ up to \emph{isometry} (any distance-preserving transformation), but is large with a quadratic size in the number $m$ of residues and fails to distinguish mirror images.
Adding a sign of orientation creates discontinuity for polygonal chains that are almost mirror-symmetric. 
\medskip

Problem~\ref{pro:map} formalizes the practically important conditions that were not all previously proved for earlier descriptors of proteins, see 
section~\ref{sec:review}.

\begin{pro}[mapping the Backbone Rigid Space] 
\label{pro:map}
For $m\geq 1$, design a map $I:\bris_m\to\R^k$  for some $k$ satisfying the conditions below.
\medskip

\noindent
(a) \textbf{Completeness}: 
any backbones $S,Q\subset\R^3$ are rigidly equivalent if and only if $I(S)=I(Q)$, i.e. 
$I$ has \emph{no false negatives} and \emph{no false positives}.
\medskip

\noindent
(b) \textbf{Reconstruction}:
any protein backbone $S\subset\R^3$ can be reconstructed from its invariant value $I(S)$ uniquely under rigid motion.
\medskip

\noindent
(c) \textbf{Lipschitz continuity}: 
there is a distance $d$ satisfying the metric axioms 
(1) $d(a,b)=0$ if and only if $a=b$;
(2) $d(a,b)=d(b,a)$;
(3) triangle inequality $d(a,b)+d(b,c)\geq d(a,c)$ for all invariant values $a,b,c$;
and a constant $\la$ such that, for any $\ep>0$, if $Q$ is obtained from $S$ by perturbing every atom up to Euclidean distance 
$\ep$, then $d(I(S),I(Q))\leq\la\ep$.
\medskip

\noindent
(d) \textbf{Atom matching} (inverse Lipschitz continuity): 
there is a constant $\mu$ such that, for any backbones $S,Q$ 
with $\de=d(I(S),I(Q))$, all their atoms can be matched up to a distance $\mu\de$ by a rigid motion. 
\medskip

\noindent
(e) \textbf{Respecting subchains}: 
for any subchain of residues $R_{i}\cup\dots\cup R_{i+j}$ in a backbone $S$,
the invariant $I(R_{i}\cup\dots\cup R_{i+j})$ can be obtained from $I(S)$ in linear time $O(j)$ with respect to the length of the subchain.
\medskip

\noindent
(f) \textbf{Linear time}: the invariant $I$, the metric $d$, a reconstruction in (b), and a rigid motion in (d) can be computed in time $O(m)$ for 
$m$ residues.
\end{pro}

The completeness in \ref{pro:map}(a) means that $I$ is the strongest possible invariant and hence \emph{distinguishes all} protein backbones that cannot be exactly matched by rigid motion.
The reconstruction in \ref{pro:map}(b) is more practical because $I$ may not allow an efficiently computable inverse map $I^{-1}$ from an invariant value $I(S)$ to a backbone $S\subset\R^3$.
The metric axioms for a distance $d$ in \ref{pro:map}(c) are essential because if the triangle axiom fails with any positive error, results of clustering can be made arbitrary \cite{rass2024metricizing}.
\medskip

The continuity in \ref{pro:map}(c) fails for invariants based on principal directions that can discontinuously change (or become ill-defined) in degenerate cases when eigenvalues become equal.  
The atom matching in \ref{pro:map}(d) says that, after finding a rigid motion $f$ in $\R^3$, any atom $p\in S$ (say, $\al$-carbon $A_i(S)$ in the $i$-th residue) has Euclidean distance at most $\mu\de$ to the corresponding atom $q\in f(Q)$, also the $\al$-carbon atom $A_i(Q)$ in the $i$-th residue of $Q$.
\medskip

Conditions~\ref{pro:map}(c,d) guarantee the Lipschitz continuity of $I$ and its inverse on the image 
$I(\bris_m)\subset\R^k$.
New condition~\ref{pro:map}(e) is important for identifying secondary structures, which are subchains in full backbones.
\medskip

The linear time in \ref{pro:map}(f) makes all previous conditions practically useful because even the distance matrix needs $O(m^2)$ time and space, substantially slower than linear time $O(m)$ for thousands of residues.
\medskip

\textbf{The key contribution} is the \emph{Backbone Rigid Invariant} $\bri$, a map $\bris_m\to\R^{9m-6}$ that 
solves Problem~\ref{pro:map}. 
Conditions~\ref{pro:map}(d,e) are stated for the first time to the best of our knowledge.
Section~\ref{sec:duplicates} will describe how $\bri$ detected thousands of unexpected geometric duplicates in the PDB, some of which require corrections, already confirmed by their authors.
\medskip

The numerical components of $\bri$ play the role of geographic-style coordinates on the space $\bris_m$, where any protein chain has a uniquely defined location.
Sections~\ref{sec:invariant} and~\ref{sec:barcode} will discuss 2D projections of the full Backbone Rigid Space $\bris=\bigcup\limits_{m\geq 2}\bris_m$ and reveal substantial variability of traditional invariants in the PDB such as bond angles and lengths, which were previously expected to have fixed values for all proteins.

\section{Past work on similarities of proteins}
\label{sec:review}


In the more general context of crystal structures, a canonical description in a reduced unit cell \cite{niggli1928krystallographische} can be achieved by the program TYPIX \cite{parthe2013typix} for inorganic compounds and ACHESYM \cite{kowiel2014achesym} for macromolecular crystals.
Such conventional settings can be considered a complete invariant in the sense of condition~(\ref{pro:map}a).
However, a reduced cell discontinuously changes under almost any perturbation of atoms, which has been known at least since 1965 \cite[p.~80]{lawton1965reduced} and was resolved only for generic crystals \cite{widdowson2022resolving}.
\medskip

The majority of past approaches to quantify protein similarity use a geometric alignment by finding an optimal rigid motion that makes a given structure as close as possible to a template (reference) protein backbone.
\medskip

The widely used TM-score \cite{zhang2004scoring}  $\TM=\max\left\{ \frac{1}{L_N}\sum\limits_{i=1}^{L_T} \frac{1}{1+(d_i/d_0)^2} \right\}\in[0,1]$ is maximized over all spatial alignments of two backbones, where $\frac{d_i}{d_0}$ is a normalized distance between aligned atoms, $L_T$ is the length (number of atoms) of a template backbone, $L_N$ is the length of a given backbone.
Since any identical backbones (with all equal coordinates $x,y,z$) have TM-score 1, one way to convert this similarity into a distance is to set $\TMD=1-\TM$ so that $\TMD(S,S)=0$ for any structure $S$.
This and many other conversions such as $d=-\log(\TM)$ fail the triangle axiom of a metric in \ref{pro:map}(c).
Indeed, let $L_T=L_N=1$ and $d_i/d_0$ be pairwise distances $\frac{1}{2},\frac{1}{3},\frac{1}{4}$ between 3 atoms, which satisfy the triangle axiom ($\frac{1}{3}+\frac{1}{4}\geq\frac{1}{2}$).
Then $\TMD=1-\TM$ takes the values $\frac{1}{5},\frac{1}{10},\frac{1}{17}$ and fails this axiom ($\frac{1}{10}+\frac{1}{17}<\frac{1}{5}$), and so does $d=-\log(\TM)$ with (approximate) values $0.22,0.11,0.06$. 
\medskip

If the triangle axiom fails with any additive error, results of the clustering algorithms $k$-means and DBSCAN can be arbitrarily pre-determined \cite{rass2024metricizing}.
The authors of another similarity LDDT (Local Distance Difference Test) concluded in \cite[p.~2728]{mariani2013lddt} that ``One disadvantage of the LDDT score is that it does not fulfill the mathematical criteria to be a metric. However, the same is true for most scores''.
One metric satisfying all axioms is the Root Mean Square Deviation (RMSD) between optimally aligned ordered atoms \cite{carugo2003root}.
This RMSD is slow to compute for all-vs-all comparisons in the PDB.
As a result, many pairs with $\RMSD=0$ remained unnoticed.
\medskip

If the order of atoms is ignored, the optimal alignment is NP-complete (provably too slow) \cite{lathrop1994protein}.
Applying random rotations \cite{draizen2024deep} creates many more structures that look different but should be considered rigidly equivalent.
This `data augmentation' makes the classification even harder.
\medskip

The PDB recently implemented a structural superposition \cite{structural} of protein backbones by computing the score equal to the sum of absolute values in the upper triangle of the distance-difference matrix (DDM) for the distance matrices between all $\al$-carbon atoms $C_\al$.
The description in \cite{structural} adds that ``to account for possible gaps in the DDMs, caused by a lack of residue coordinates, these scores are multiplied by a scalar between 0-1, where 1 represents the absence of any gaps ...  low scores represent chains with high structural similarity.''
This scaling by values less than 1 likely affects the triangle axiom, which needs checking in the light of the recent reviews \cite{terwilliger2024alphafold,jones2022impact,moore2022protein} of protein folding prediction \cite{jumper2021highly,leman2020macromolecular,mirdita2022colabfold}.
\medskip

More importantly, to efficiently navigate in the protein universe, in addition to distances, we need a map showing all known structures and also under-explored regions, where new proteins can be discovered.
Such a geographic-style map needs a complete invertible and bi-continuous invariant $I$ like the pair of latitude and longitude coordinates on Earth.
\medskip

Protein backbones are traditionally represented  by \emph{torsion} (dihedral) angles $\ph_i,\psi_i$ visualized in Ramachandran plots \cite{ramachandran1968conformation}.
For a general polygonal line on points $S\subset\R^3$, the sequence $\{\phi_i,\psi_i\}$ is invariant under rigid motion but incomplete.
Indeed, for any successive points $p_i,p_{i+1}\in S$, we can shift all points $p_{i+1},\dots,p_{m}$ by a vector $t(\vec p_{i+1}-\vec p_i)$ for any $t\neq 0$, which changes the overall rigid shape of $S$ but keeps all relative angles between any straight segments and planes through successive points.
\medskip

For protein backbones, even if all bond lengths and angles are fixed at ideal values, all torsion angles still should be ordered according to given residues to completely determine the rigid class of a backbone.
Even if we keep all torsion angles in order, three invariants per residue cannot uniquely determine a rigid backbone having 3 atoms with 9 coordinates per residue in $\R^3$.
AlphaFold2 \cite{jumper2021highly} used 6 parameters per residue to define a rigid transformation on every $i$-th triplet (\emph{residue triangle}) on the main atoms $N_i,A_i,C_i$ to the next $(i+1)$-st residue triangle.
However, the analysis in section~\ref{sec:invariant} will show that rigid shapes of residue triangles substantially vary across the PDB.
Our paper strengthens the past approach by defining 9 invariants per each of $m$ residues, which gives $9m-6$ invariants in total after subtracting 6 parameters of a global rigid motion in $\R^3$.
\medskip

If we consider a backbone $S$ of $3m$ ordered atoms modulo isometry including reflections, the easier complete invariant known since 1935 \cite{schoenberg1935remarks} is the $3m\times 3m$ matrix $D(S)$ of all pairwise distances whose entry $D_{ij}(S)$ is the Euclidean distance between the $i$-th and $j$-th points of $S$.
Any backbone $S$ can be reconstructed from $D(S)$ or, equivalently, from the Gram matrix of scalar products as in \cite[Theorem~1]{dekster1987edge}, uniquely up to isometry in $\R^3$.
The matrix $D(S)$ satisfies almost all conditions of Problem~\ref{pro:map} apart from the linear time/size requirement, which is essential for large proteins. 
\medskip

If a protein is considered a cloud of unordered atoms (ignoring the order along a backbone), such clouds of different sizes can be visualized by eigenvalue invariants (or moments of inertia) characterizing the elongation of the cloud along its principal directions.
In 1996, probably the first map of all 4K entries in the PDB appeared in \cite[Fig.~5]{holm1996mapping} based on the two largest eigenvalues, see the recent updates in \cite[Fig.~2]{wirth2013protein} and PDB-Explorer \cite{jin2015pdb}. 
\medskip

In 2020, Holm called for faster visualization of the protein space \cite{holm2020dali}: ``It would be nice to restore the ability to move a lens across fold space in real-time ... this ability was based on pre-computed all-against-all structural similarities, which is not manageable with current data''.  
\medskip

In 1977, Kendall \cite{kendall1977diffusion} started to study configuration spaces of ordered points modulo rigid motion in $\R^n$ under the name of \emph{size-and-shape spaces} \cite{kendall2009shape}.
If we consider sequences equivalent also under uniform scaling, the smaller \emph{shape space} $\Si_2^m$ of $m$ ordered points in $\R^2$ can be described as a complex projective space $\C P^{m-1}$ due to the group $\SO(2)$ being identified with the unit circle in the complex space $\C^1=\R^2$. 
However, there is no easy description of the space $\Si_3^m$ of $m$-point sequences in $\R^3$, which has no multiplicative group structure similar to $\R^2=\C^1$.
This algebraic obstacle prevented a simple solution to Problem~\ref{pro:map} in dimension $n=3$.

\section{The backbone rigid invariant ($\bri$)}
\label{sec:invariant}

We start with the simpler \emph{triangular invariant} that describes the rigid shape of each residue triangle $\triangle N_i A_i C_i$ on three main atoms per each of $m$ residues: nitrogen $N_i$, $\alpha$-carbon $A_i$, and carbonyl carbon $C_i$, for $i=1,\dots,m$, see Fig.~\ref{fig:backbones}~(middle).
For any points $A,B\in\R^3$, let $|\ve{AB}|$ be the Euclidean length of the vector $\ve{AB}$ from $A$ to $B$.
We denote vectors by $\vb*{u}\in\R^3$, their \emph{scalar} and \emph{vector} products by $\vb*{u}\cdot\vb*{v}$ and $\vb*{u}\times\vb*{v}$, respectively. 

\begin{dfn}[triangular invariant $\trin$]
\label{dfn:trin}
Let a backbone $S\subset\R^3$ have $3m$ ordered atoms $N_i$, $A_i$, $C_i$, $i=1,\dots,m$.
In the plane of $\triangle N_i A_i C_i$, for the 2D basis obtained by Gaussian orthogonalization of $\ve{A_i N_i},\ve{A_i C_i}$, the vector $\ve{A_i N_i}$ has the coordinates $x(A_i N_i)=|\ve{A_i N_i}|$,  $y(A_i N_i)=0$, while $\ve{A_i C_i}$ has $x(A_i C_i)=\dfrac{\ve{A_i C_i}\cdot\ve{A_i N_i}}{|\ve{A_i N_i}|}$ and
$y(A_i C_i)=\Big|\ve{A_i C_i} - x(A_i C_i)\dfrac{\ve{A_i N_i}}{|\ve{A_i N_i}|}\Big|$.
The \emph{triangular invariant} $\trin(S)$ is the $m\times 3$ matrix whose $i$-th row consists of the coordinates $x(A_i N_i),x(A_i C_i),y(A_i C_i)$ for $i=1,\dots,m$.
\end{dfn}

The $i$-th row of $\trin(S)$ uniquely determines the shape of $\triangle N_i A_i C_i$.
Many past approaches including AlphaFold2 \cite{jumper2021highly} assumed that all these residue triangles are rigidly equivalent.
To test this assumption on the PDB, we filter out unsuitable chains as follows.
On May 4, 2024, the PDB had 213,191 entries with 1,091,420 chains. 
Protocol~\ref{prot:cleanPDB} below produced $104,688\approx 49\%$ entries with $707410\approx 65\%$ chains in 4 hours 48 min 11 sec.
All experiments were run on 
CPU Core i7-11700 @2.50GHz RAM 32Gb.

\begin{prot}[selecting a subset of 707K+ chains in the PDB]
\label{prot:cleanPDB}
The PDB was filtered by removing the following entries and individual chains. \\
(1) 4513 non-proteins (the entity is labeled as `not a protein'). \\
(2) 178153 disordered chains, where some atoms have occupancies $<1$. \\
(3) 201648 chains with residues having non-consecutive indices. \\
(4) 9941 incomplete chains missing one of the main atoms $N_i,A_i,C_i$. \\
(5) 4364 chains with non-standard amino acids.
\end{prot}
 
\begin{figure}[ht!]
\includegraphics[width=0.325\textwidth]
{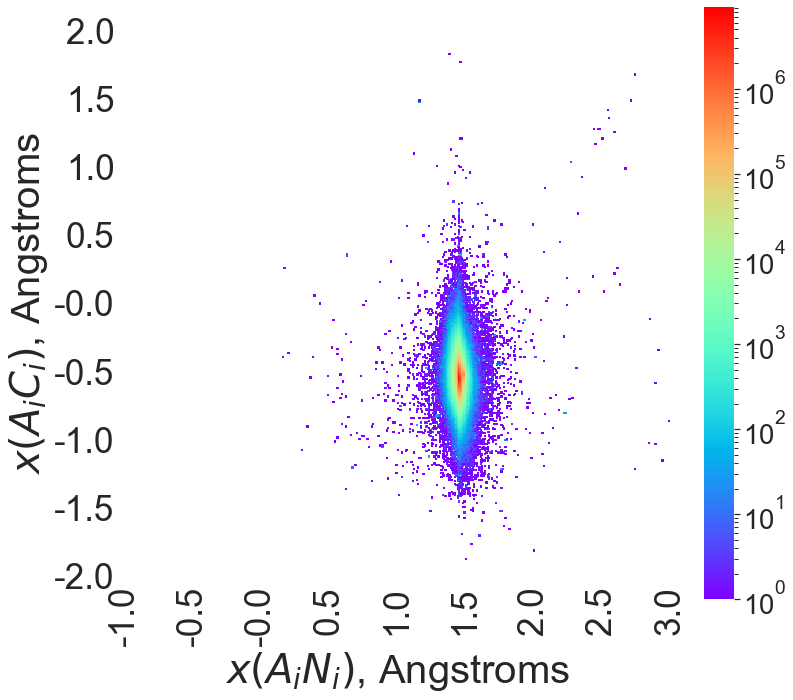}
\includegraphics[width=0.325\textwidth]
{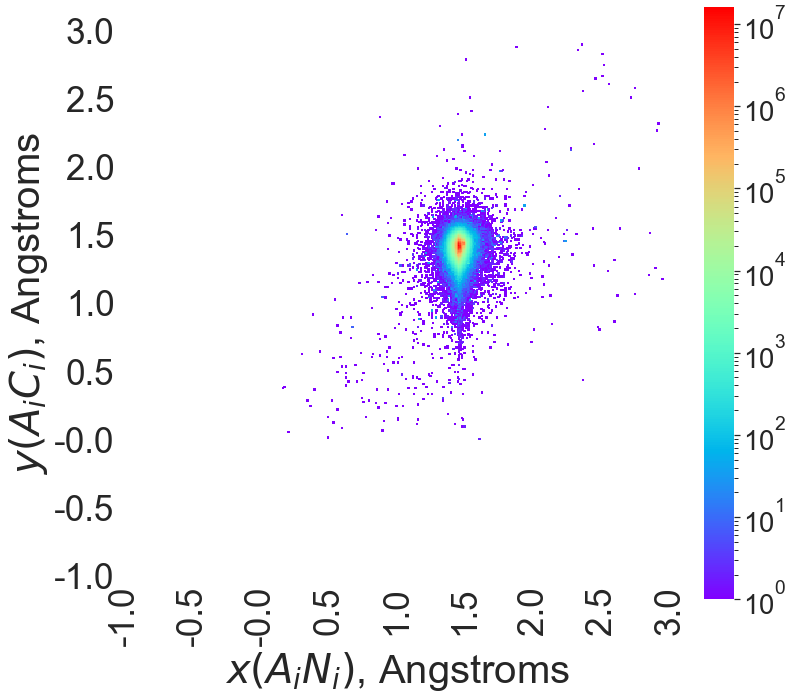}
\includegraphics[width=0.325\textwidth]
{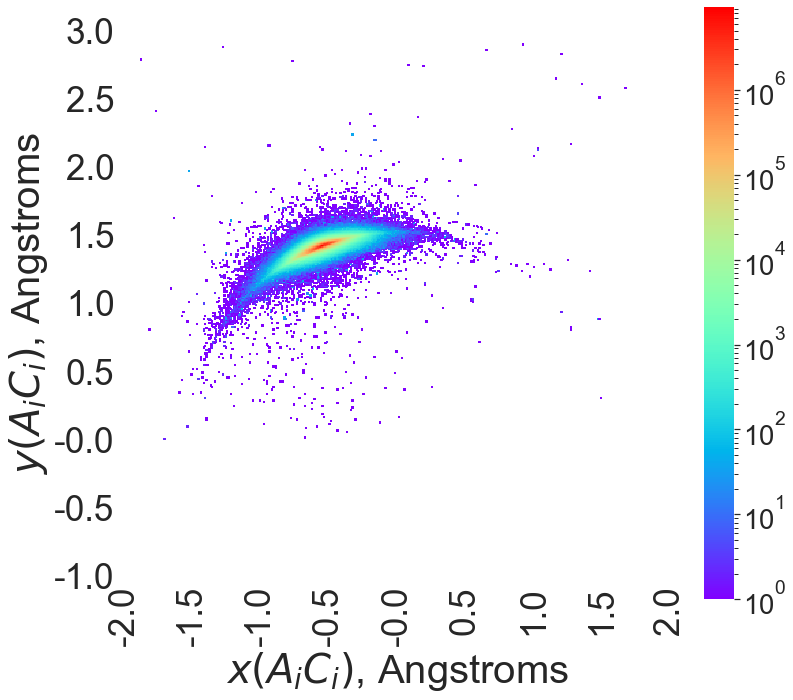}
\bigskip

\includegraphics[width=0.325\textwidth]{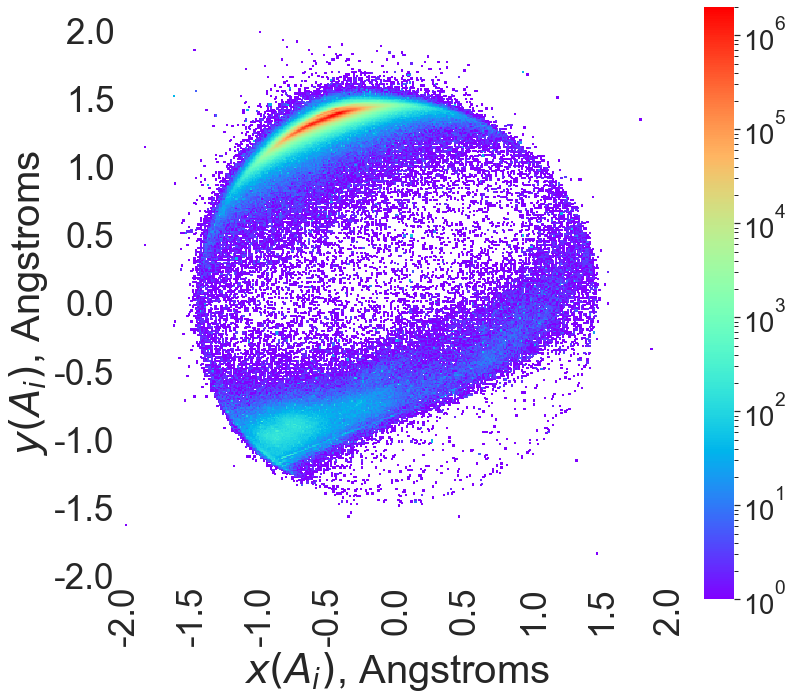}
\includegraphics[width=0.325\textwidth]{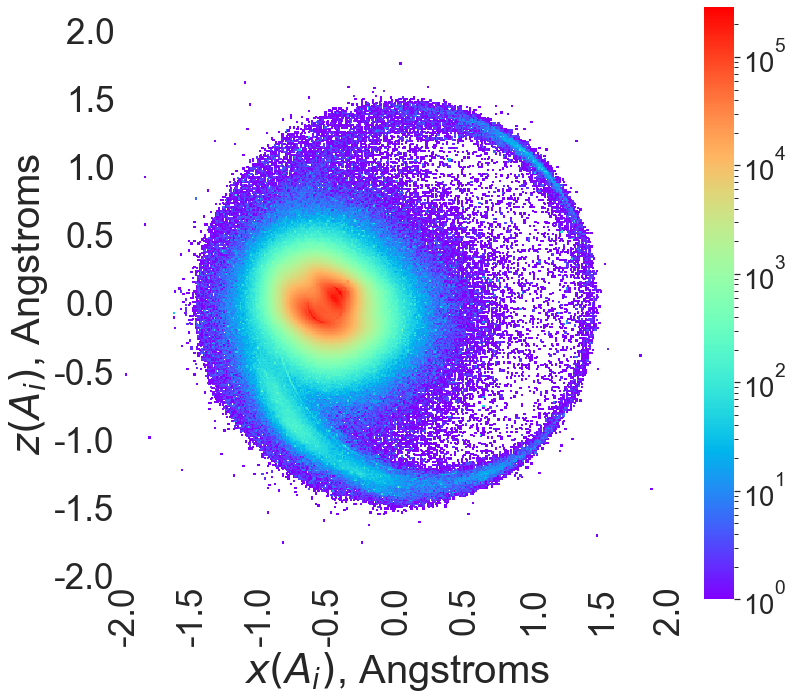}
\includegraphics[width=0.325\textwidth]{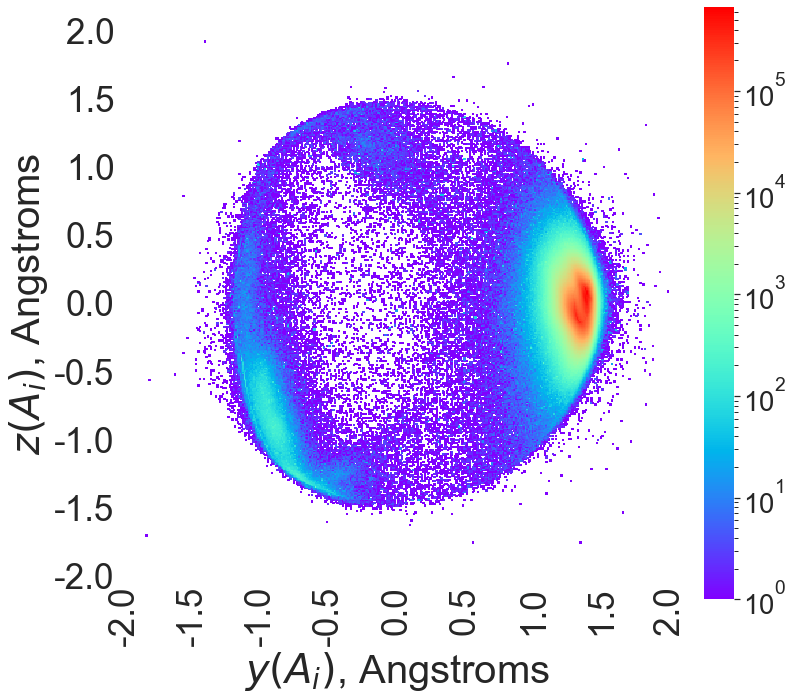}
\bigskip

\includegraphics[width=0.325\textwidth]{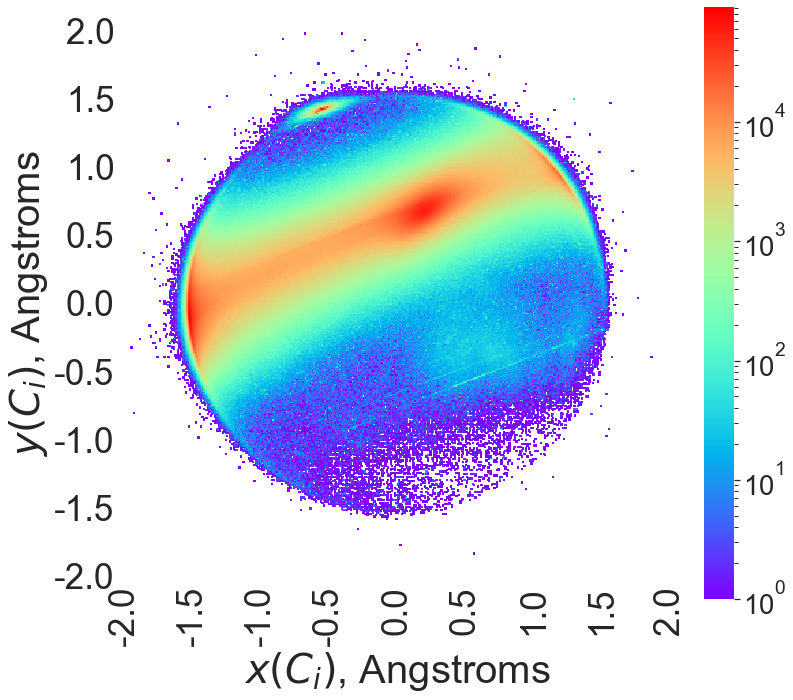}
\includegraphics[width=0.325\textwidth]{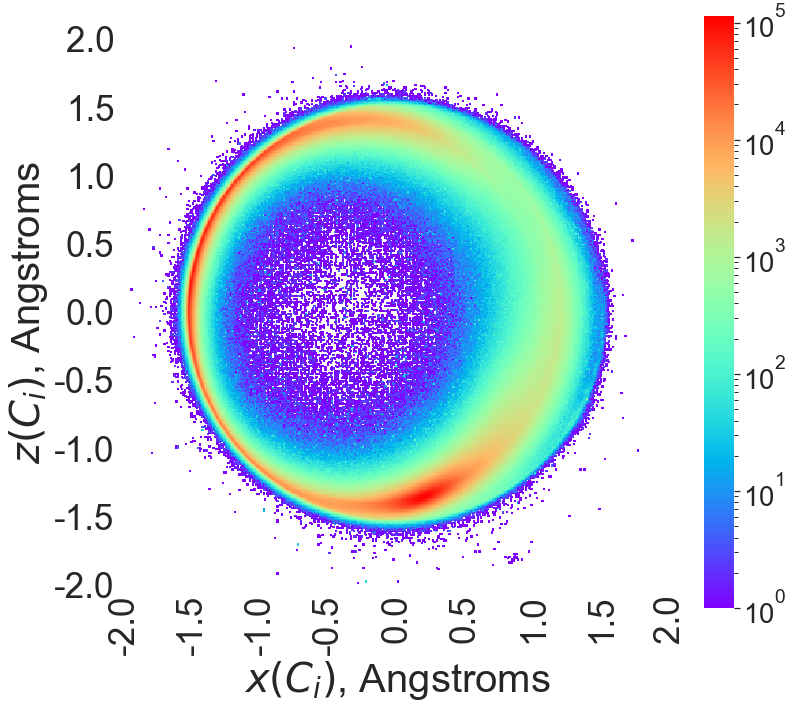}
\includegraphics[width=0.325\textwidth]{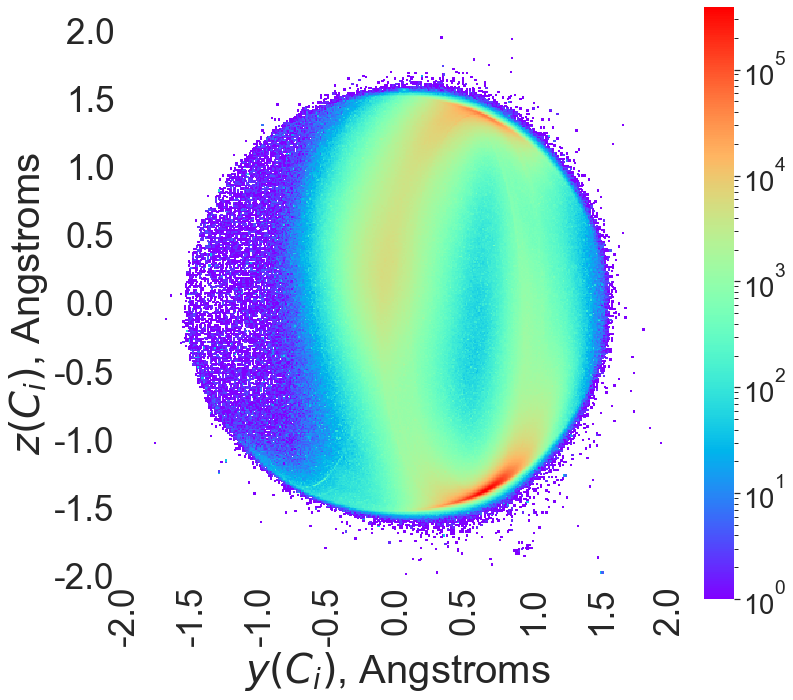}
\bigskip

\includegraphics[width=0.325\textwidth]{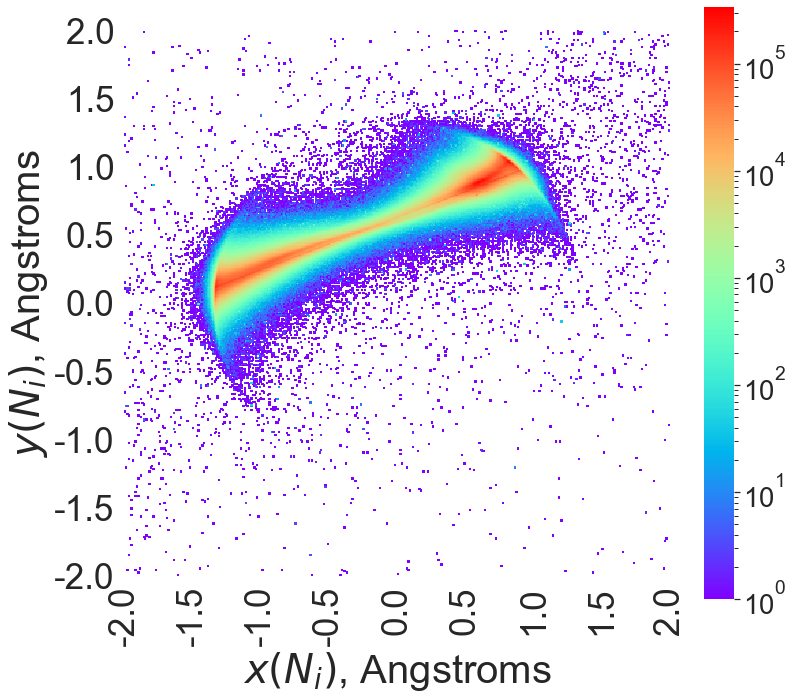}
\includegraphics[width=0.325\textwidth]{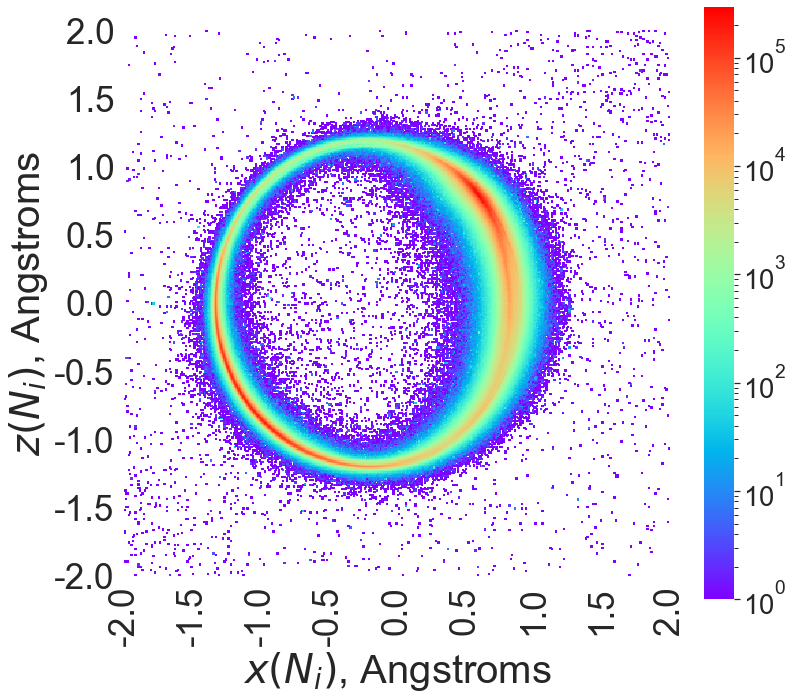}
\includegraphics[width=0.325\textwidth]{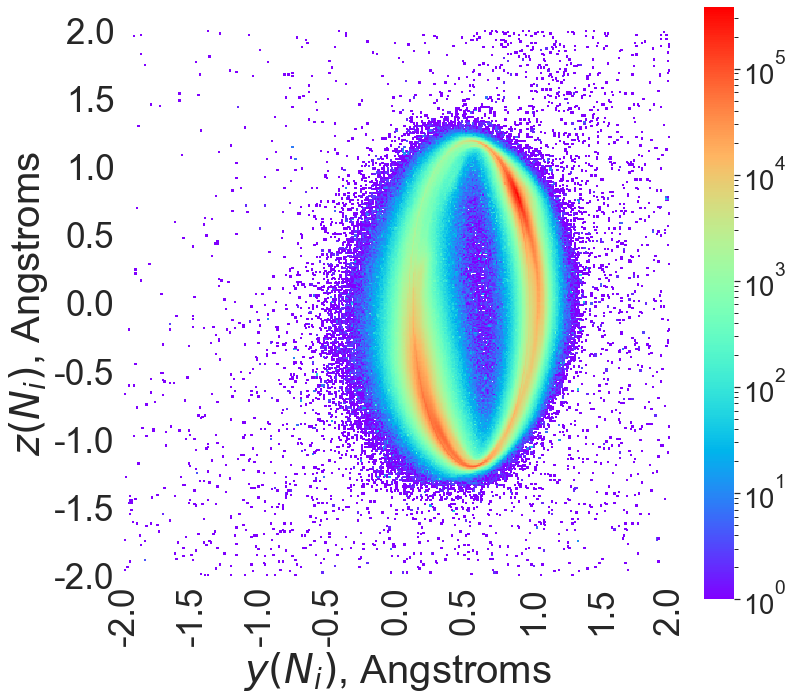}

\caption{
Heatmaps of the invariants $\trin$ and $\bri$ from Definitions~\ref{dfn:trin} and~\ref{dfn:bri} for all 110M+ residues in the 707K+ chains obtained by Protocol~\ref{prot:cleanPDB}.
The color indicates the number of residues whose pair of invariants is discretized to each pixel.
}
\label{fig:PDB707K_BRI_heatmaps_log}
\end{figure}

\begin{exa}[variability of residue triangles]
\label{exa:triangles}
Fig.~\ref{fig:PDB707K_BRI_heatmaps_log}~(row 1) shows
the heatmaps of the invariants $x(A_i N_i),x(A_i C_i),y(A_i C_i)$  on a logarithmic scale from Definition~\ref{dfn:trin} across all 110+ million residues from the 707K+ cleaned backbones obtained by Protocol~\ref{prot:cleanPDB}.  
Though standard deviations of these invariants are about $0.01\angstrom$, the maximum deviations of $x(A_i N_i),x(A_i C_i),y(A_i C_i)$ have high values of $1.2,1.7,2.7\angstrom$, respectively.
\medskip

Table~\ref{tab:haemoglobin_invariants} below shows the coordinates of TRIN and BRI (see Definition~\ref{dfn:bri}) for the two hemoglobin chains A in proteins 2hhb and 1hho, which are shown in Fig.~\ref{fig:hemoglobins}~(top middle) and discussed in Example \ref{exa:hemoglobins}. 

\begin{table}[h!]
\caption{Coordinates of $\trin$ and $\bri$ for the first 3 residues of the chains A in 2hhb (top) and 1hho (bottom) with their means. 
}

\centering
\resizebox{\textwidth}{!}{%
    \begin{tabular}{l|llllllllllll}
    Res & $x(AN)$ & $x(AC)$ & $y(AC)$ & $x(N)$ & $y(N)$ & $z(N)$ & $x(A)$ & $y(A)$ & $z(A)$ & $x(C)$ & $y(C)$ & $z(C)$\\
    \hline
    VAL&1.45&-0.54&1.44&1.45&0&0&0&0&0&-0.54&1.44&0\\
    LEU&1.47&-0.50&1.47&-0.91&0.25&-0.90&-0.64&1.32&0.02&-1.10&0.01&1.10\\
    SER&1.47&-0.48&1.45&-0.77&0.36&-0.98&-0.66&1.31&-0.05&-1.11&0.02&1.06\\
    mean &1.47&-0.55&1.43&0.52&0.84&0.46&-0.48&1.38&0.05&0.01&0.65&-1.01
\vspace{2em}
    \end{tabular}}

\centering
\resizebox{\textwidth}{!}{%
    \begin{tabular}{l|llllllllllll}
    Res & $x(AN)$ & $x(AC)$ & $y(AC)$ & $x(N)$ & $y(N)$ & $z(N)$ & $x(A)$ & $y(A)$ & $z(A)$ & $x(C)$ & $y(C)$ & $z(C)$\\
    \hline
    VAL&1.48&-0.51&1.46&1.48&0.00&0.00&0.00&0.00&0.00&-0.51&1.46&0.00\\
    LEU&1.49&-0.55&1.42&-0.14&0.66&1.16&-0.69&1.31&0.19&-1.51&-0.16&-0.03\\
    SER&1.44&-0.41&1.44&-0.63&0.27&-1.10&-0.36&1.36&-0.30&-1.43&0.14&0.40\\
    mean &1.47&-0.53&1.43&0.56&0.81&0.44&-0.43&1.38&0.06&0.04&0.65&-1.02
    \end{tabular}}
\label{tab:haemoglobin_invariants}
\end{table}
\end{exa}  

To guarantee new condition~\ref{pro:map}(e) respecting subchains,
Definition~\ref{dfn:bri} will represent atoms $N_{i+1},A_{i+1},C_{i+1}$ in a basis of the  previous $i$-th residue.
The first residue needs only three invariants from Definition~\ref{dfn:trin} to determine the rigid shape of $\triangle N_1 A_1 C_1$ in $\R^3$. 
Due to cleaning in Protocol~\ref{prot:cleanPDB}, all consecutive atoms along any backbone have distances $d\geq 0.01\angstrom$ and all angles in any residue triangle $\triangle N_i A_i C_i$ are at least $3^\circ$, which makes the bases of all residue triangles well-defined in Definition~\ref{dfn:bri} below.

\clearpage

\begin{dfn}[backbone rigid invariant $\bri(S)$ of a protein backbone $S$]
\label{dfn:bri}
In the notations of Definition~\ref{dfn:trin}, 
define the orthonormal basis vectors 
$\vb*{u}_i=\dfrac{\ve{A_i N_i}}{|\ve{A_i N_i}|}$, 
$\vb*{v}_i=\dfrac{\vb*{h}_i}{|\vb*{h}_i|}$ for $\vb*{h}_i=\ve{A_i C_i}-b_i\ve{A_i N_i}$, 
$b_i=\dfrac{\ve{A_i C_i}\cdot\ve{A_i N_i}}{|\ve{A_i N_i}|^2}$, and $\vb*{w}_i=\vb*{u}_i\times\vb*{v}_i$.
The \emph{backbone rigid invariant} $\bri(S)$ is the $m\times 9$ matrix whose $i$-th row for $i=2,\dots,m$ contains the coefficients $x,y,z$ of the vectors $\ve{C_{i-1}N_i}$, $\ve{N_i A_i}$, $\ve{A_i C_i}$ in the basis $\vb*{u}_{i-1},\vb*{v}_{i-1},\vb*{w}_{i-1}$.
So the nine columns of $\bri(S)$ contain the coordinates $x(N_i),y(N_i),z(N_i)$ of 
$\ve{C_{i-1}N_i}$, 
followed by the six coordinates $x(A_i),\dots,z(C_i)$. 
For $i=1$, the first row of $\bri(S)$ has only three non-zero coordinates $x(N_1)=x(A_1 N_1)$, $x(C_1)=x(A_1 C_1)$, $y(C_1)=y(A_1 C_1)$ from the first row of 
$\trin(S)$ in Definition~\ref{dfn:trin}. 
\end{dfn}

Fig.~\ref{fig:PDB707K_BRI_heatmaps_log} shows heatmaps of the PDB cleaned by Protocol~\ref{prot:cleanPDB}.
We mapped each of 110+ million residues across all 707+ thousand chains to a pair of coordinates $(x,y)$ from the invariants $\trin$ and $\bri$.
When many points $(x,y)$ were discretized to a single pixel, its color reflects the number of such points on a logarithmic scale in the color bars of all heatmaps.
\medskip

For a backbone of $m$ residues, the first row of the $m\times 9$ matrix $\bri(S)$ contains only three non-zero coordinates.
Hence the matrix $\bri(S)$ can be considered a vector of length $9(m-1)+3=9m-6$.
The simplest metric on backbone rigid invariants as vectors in $\R^{9m-6}$ is $L_\infty$ equal to the maximum absolute difference between all corresponding coordinates.
\medskip

A small value $\de$ of $L_\infty(\bri(S),\bri(Q))$ guarantees by Theorem~\ref{thm:inverse} that backbones $S,Q$ are closely matched by rigid motion.
Another metric such as Euclidean distance or its normalization by the chain length has no such guarantees and can be small even for a few outliers that can affect the rigid shape and hence functional properties of a protein. 
Theorem~\ref{thm:motion} proves conditions~\ref{pro:map}(a,b,c,e,f) in Problem~\ref{pro:map} for the invariant $\bri(S)$.

\begin{thm}[completeness, reconstruction, and subchains]
\label{thm:motion}
\textbf{(a)}
Under any rigid motion in $\R^3$, the matrix $\trin(S)$ in Definition~\ref{dfn:trin} is invariant, $\bri(S)$ in Definition~\ref{dfn:bri} is a complete invariant, so any backbones $S,Q\subset\R^3$ are matched by rigid motion if and only if $\bri(S)=\bri(Q)$.
\medskip

\noindent
\textbf{(b)}
The invariant $\bri(S)$, metric $L_\infty$ between invariants, and a reconstruction of $S\subset\R^3$ from $\bri(S)$ can be computed in time $O(m)$. 
\medskip

\noindent
\textbf{(c)}
Let $Q$ be a subchain of $j$ consecutive residues in a backbone $S\subset\R^3$.
If $Q$ includes the first residue of $S$, then $\bri(Q)$ consists of the first $j$ rows of $\bri(S)$.
If $Q$ starts from the $i$-th residue of $S$ for $i>1$, the rows $2,\dots,j$ of $\bri(Q)$ coincide with the rows $i+1,\dots,i+j-1$ of $\bri(S)$, and the 1st row of $\bri(Q)$ is computed from the $i$-th row of $\bri(S)$ in a constant time.
Hence $\bri(Q)$ is computed from $\bri(S)$ in time $O(j)$. 
\end{thm}
\begin{proof}[Proof of Theorem~\ref{thm:motion}]
\textbf{(a,b)}
The formulae of the basis vectors in Definition~\ref{dfn:bri}
guarantee that all vectors have unit length $|\vb*{u}_i|=|\vb*{v}_i|=|\vb*{w}_i|=1$ and are orthogonal to each other due to $\vb*{u}_i\cdot\vb*{v}_i=\vb*{v}_i\cdot\vb*{w}_i=\vb*{w}_i\cdot\vb*{u}_i=0$.
Any rigid motion $f$ acting on a backbone $S\subset\R^3$ has the form $f(p)=\vec v+R(\vec p)$ for a fixed vector $\vec v\in\R^n$, an orthogonal map $R\in\Or(\R^3)$, and any $p\in\R^3$.
Then $f$ maps every orthonormal basis $\vb*{u}_i,\vb*{v}_i,\vb*{w}_i$ with the origin at a point $A_i\in\R^3$ to another orthonormal basis $R(\vb*{u}_i)$, $R(\vb*{v}_i)$, $R(\vb*{w}_i)$ at the new origin $f(A_i)$.
Hence the image of any vector $\ve{A_i P_i}=x\vb*{u}_i +y\vb*{v}_i +z\vb*{w}_i$ under $f$ has the same coordinates in the rigidly transformed basis: $f(\ve{A_i P_i})=R(\ve{A_i P_i})= xR(\vb*{u}_i) +yR(\vb*{v}_i) +zR(\vb*{w}_i)$, so $\bri(S)=\bri(f(S))$.
\medskip

For any residue having a fixed index $i$, Definition~\ref{dfn:bri} needs only a constant time $O(1)$ to compute the basis vectors and coordinates of $\ve{C_{i-1}N_i}$ in the basis of the previous residue.
The total time for computing the $m\times 9$ matrix $\bri(S)$ is $O(m)$.
The metric $L_\infty$ has a linear time in the size $9m$. 
\medskip

The completeness will follow by showing that any backbone $S\subset\R^3$ can be efficiently reconstructed from $\bri(S)$, uniquely after fixing the first residue whose shape is determined by the three non-zero values in the first row of $\bri(S)$.
In the first residue, the $\al$-carbon $A_1$ can be moved to the origin $0\in\R^3$ by translation. 
Using $x(N_1)=|\ve{A_1 N_1}|$, the $N$-terminal atom $N_1$ can be fixed in the positive $x$-axis by an orthogonal map from $\SO(3)$. 
A suitable rotation around the $x$-axis can move $C_1$ to the upper $xy$-plane.
All these transformations preserve the lengths and scalar products.
The final position of $C_1$ is uniquely determined by $x(C_1),y(C_1)$ in Definition~\ref{dfn:bri}.
\medskip

After fixing $\triangle N_1,A_1,C_1$, it remains to prove that any other atom of $S$ is uniquely determined by its $x,y,z$ coordinates in $\bri(S)$.
Indeed, $N_2$ is obtained from $C_1$ by adding $\ve{C_1N_2}$, whose coordinates are the first three elements in the 2nd row of $\bri(S)$.
Then $A_2$ is obtained from $N_1$ by adding $\ve{N_2A_2}$,
whose coordinates are the next three elements in the 2nd row of $\bri(S)$.
Then $C_2$ is obtained from $A_2$ by adding $\ve{A_2C_2}$ and so on.
\medskip

\noindent
\textbf{(c)}
Since the complete invariant $\bri(S)$ of a backbone $S$ is locally defined by determining any $i$-th residue triangle $\triangle N_i A_i C_i$ in the basis of the previous $(i-1)$-st triangle, all rows of the matrix $\bri(Q)$ for any subchain $Q$ in the full backbone $S$ coincide with the corresponding rows of $\bri(S)$.
\medskip

The only exception is the first row if $Q$ starts from the $i$-th residue of $S$ for $i>1$.
In this case, the three non-zero invariants in the first row of $Q$ can be obtained from the $i$-th row of $\trin(S)$ whose values are expressed in terms of the vectors $\ve{N_i A_i}$ and $\ve{A_i C_i}$ in Definition~\ref{dfn:trin}.
This computation needs only a constant time independent of $j$ because the coordinates of the vectors $\ve{A_i N_i}$ and $\ve{A_i C_i}$ are given in the $i$-th row of $\bri(S)$.
\end{proof}

\begin{cor}[completeness under isometry]
\label{cor:isometry}
Any mirror image $\bar S$ of a backbone $S\subset\R^3$ has the invariant $\ov{\bri}(S):=\bri(\bar S)$ obtained by reversing the signs in all $z$-columns of $\bri(S)$.
The unordered pair of $\bri(S)$ and $\ov{\bri}(S)$ is complete under isometry.
\end{cor}

\begin{proof}[Proof of Corollary~\ref{cor:isometry}]
To prove that $\ov{\bri}(S):=\bri(\bar S)$ is obtained from $\bri(S)$ by reversing the signs in all $z$-columns of $\bri(S)$, consider the main atoms $N_i,A_i,C_i$ in the $i$-th residue of $S$ for any $i=2,\dots,m$.
The mirror image $\bar S$ has the corresponding atoms $\bar N_i,\bar A_i,\bar C_i$.
There is a rigid motion $f$ in $\R^3$ that matches these atoms so that $N_i=f(\bar N_i)$, $A_i=f(\bar A_i)$, $C_i=f(\bar C_i)$, and $f(\bar S)$ is obtained from $S$ by the reflection $g$ in the plane of the residue triangle $\triangle N_i A_i C_i$.
This reflection $g$ preserves the basis vectors $\vb*{u}_i,\vb*{v}_i,\vb*{w}_i$ from Definition~\ref{dfn:bri} of the $i$-th residue of the backbone $S$.
\medskip

In the orthonormal basis of $u_i,v_i$, $w_i=u_i\times v_i$, the coordinates of the vector $\ve{C_i N_{i+1}}=x(N_{i+1})\vb*{u}_i + y(N_{i+1})\vb*{v}_i + z(N_{i+1})\vb*{w}_i$ 
determine the coordinates of the mirror image
$f(\ve{\bar C_i \bar N_{i+1})}=x(N_{i+1})\vb*{u}_i + y(N_{i+1})\vb*{v}_i - z(N_{i+1})\vb*{w}_i$, where only the sign of the coefficient of $\vb*{w}_i$ is reversed as required.
Since the index $i=2,\dots,m$ was arbitrarily chosen, it remains to notice that the first residue triangles $\triangle N_1 A_1 C_1$ and $\triangle \bar N_1 \bar A_1 \bar C_1$ can be matched by rigid motion, so all 3 non-zero invariants in the first rows of $\bri(S)$ and $\bri(\bar S)$ coincide, while all $z$-coordinates are zeros.
Finally, the unordered pair of $\bri(S)$ and $\ov{\bri}(S)$ is invariant under any rigid motion by Theorem~\ref{thm:motion}(a) and under reflection, which swaps the invariants in this pair.
By Theorem~\ref{thm:motion}(b), any of $\bri(S)$ and $\ov{\bri}(S)$ suffices to reconstruct $S$ or $\bar S$ up to rigid motion, hence $S$ up to isometry in $\R^3$.
\end{proof}

\section{Lipschitz bi-continuity of the invariant $\bri$}
\label{sec:continuity}

Theorem~\ref{thm:continuity} will prove the Lipschitz continuity of $\bri$ in condition~\ref{pro:map}(c).
For a given backbone $S$ and its perturbation $Q$, let $l_{N,A}$ and $L_{N,A}$ denote the minimum and maximum bond length between any $\al$-carbon $A_i$ and nitrogen $N_i$ in $S,Q$, respectively.
The maximum bond lengths $L_{A,C},L_{C,N}$ are similarly defined for other types of bonds.

\begin{thm}[Lipschitz continuity of $\bri$]
\label{thm:continuity}
For any $\ep>0$, let $Q$ be obtained from a backbone $S\subset\R^3$ by perturbing every atom of $S$ up to Euclidean distance $\ep$.
Let $h=\min_i|y(A_i C_i)|$ be the minimum height in triangles $\triangle N_i A_i C_i$ at $C_{i}$ for all residues in the backbones $S,Q$. 
Set $L=\max\{L_{C,N},L_{N,A},L_{A,C}\}$, $K=\dfrac{1}{l_{N,A}}+\dfrac{2}{h}\Big(1+2\dfrac{L_{A,C}}{l_{N,A}}\Big)$, and $\la=2(1+2LK)$.
Then $L_\infty(\bri(S),\bri(Q))\leq \la\ep$. 
\end{thm}
\smallskip

Theorem~\ref{thm:continuity} needs Lemmas~\ref{lem:length_difference}, \ref{lem:vec_perturbation}, \ref{lem:norm_vectors}, \ref{lem:products}, and Proposition~\ref{prop:basis}.

\begin{lem}[length difference]
\label{lem:length_difference}
Any 
$\vb*{u},\vb*{v}\in\R^n$ satisfy
$|\,|\vb*{u}|-|\vb*{v}|\,|\leq |\vb*{u}-\vb*{v}|$.
\end{lem}
\begin{proof}
The triangle inequality for the Euclidean distance implies that
$|\vb*{u}|\leq |\vb*{u}-\vb*{v}|+|\vb*{v}|$, so 
$|\vb*{u}|-|\vb*{v}|\leq |\vb*{u}-\vb*{v}|$.
Swapping the vectors, we get
$|\vb*{v}|-|\vb*{u}|\leq |\vb*{u}-\vb*{v}|$.
Combining the inequalities $\pm(|\vb*{u}|-|\vb*{v}|)\leq |\vb*{u}-\vb*{v}|$, we conclude that $|\,|\vb*{u}|-|\vb*{v}|\,|\leq |\vb*{u}-\vb*{v}|$ as required.
\end{proof}

\begin{lem}[perturbation of a vector]
\label{lem:vec_perturbation}
Let $A',B'$ be any $\ep$-perturbations of points $A,B\in\R^n$, respectively, i.e. $|A-A'|\leq\ep$, $|B-B'|\leq\ep$.
Then $|\ve{A'B'}-\ve{AB}|\leq 2\ep$.
\end{lem}
\begin{proof}
Apply the triangle inequality: \\
$|\ve{A'B'}-\ve{AB}|=|\ve{A'A}+\ve{BB'}|\leq |\ve{A'A}|+|\ve{BB'}|\leq 2\ep$.
\end{proof}

\begin{lem}[a normalized vector]
\label{lem:norm_vectors}
Let $\vb*{u}$ be a $\de$-perturbation of a vector $\vb*{v}\in\R^n$, i.e. $|\vb*{u} -\vb*{v}|\leq\de$.
Then $\left|\dfrac{\vb*{u}}{|\vb*{u}|}-\dfrac{\vb*{v}}{|\vb*{v}|}\right|\leq \dfrac{2\de}{l}$, where $l=\max\{|\vb*{u}|,|\vb*{v}|\}$.
Hence if $\vb*{u}=\ve{A N}$ and $\vb*{u}'=\ve{A' N'}$ are vectors between atoms $A_i,N_i$ and their $\ep$-perturbations, then 
$\left|\dfrac{\vb*{u}}{|\vb*{u}|}-\dfrac{\vb*{v}}{|\vb*{v}|}\right|\leq \dfrac{4\ep}{l_{N,A}}$, where $l_{N,A}$ is the minimum bond length between $N_i,A_i$.
\end{lem}
\begin{proof}
Assume that $\max\{|\vb*{u}|,|\vb*{v}|\}=|\vb*{v}|$, which we denote by $l$. 
Then
\begin{align*}
& \left|\dfrac{\vb*{u}}{|\vb*{u}|}-\dfrac{\vb*{v}}{|\vb*{v}|}\right|
=\left|\dfrac{|\vb*{v}|\vb*{u}-|\vb*{u}|\vb*{v}}{|\vb*{u}|\cdot |\vb*{v}|}\right|=
\dfrac{|\,(|\vb*{v}|-|\vb*{u}|)\vb*{u}+|\vb*{u}|(\vb*{u}-\vb*{v})\,|}{|\vb*{u}|\cdot |\vb*{v}|}\leq \\
& \dfrac{|\, |\vb*{u}|-|\vb*{v}|\,|\cdot |\vb*{u}| + |\vb*{u}|\cdot |\vb*{u}-\vb*{v}|}{|\vb*{u}|\cdot|\vb*{v}|}
=\dfrac{|\,|\vb*{u}|-|\vb*{v}|\,|+|\vb*{u}-\vb*{v}|}{|\vb*{v}|}
\leq \dfrac{2|\vb*{u}-\vb*{v}|}{|\vb*{v}|}
\leq 
\dfrac{2\de}{l},
\end{align*} 
where we used the triangle inequality, Lemma~\ref{lem:length_difference}, and $|\vb*{u} -\vb*{v}|\leq\de$.
The second inequality follows for $\de=2\ep$ from Lemma~\ref{lem:vec_perturbation} and $l_{N,C}\leq\max\{|\vb*{u}|,|\vb*{v}|\}$.
\end{proof}

\begin{lem}[products] 
\label{lem:products}
For any 
$\vb*{u},\vb*{u}',\vb*{v},\vb*{v}'\in\R^n$, 
if $|\vb*{v}'|=|\vb*{v}|=1$, 
then 
\smallskip

\noindent
\textbf{(a)} 
$|(\vb*{u}'\cdot \vb*{v}') - (\vb*{u}\cdot\vb*{v})|\leq
|\vb*{u}' - \vb*{u}|+|\vb*{u}|\cdot|\vb*{v}' - \vb*{v}|$, 
\smallskip

\noindent
\textbf{(b)} 
$|(\vb*{u}'\times \vb*{v}') - (\vb*{u}\times\vb*{v})|\leq
|\vb*{u}' - \vb*{u}|+|\vb*{u}|\cdot|\vb*{v}' - \vb*{v}|$,
\smallskip

\noindent
\textbf{(c)} 
$|(\vb*{u}'\cdot \vb*{v}')\vb*{v}' - (\vb*{u}\cdot\vb*{v})\vb*{v}|\leq
|\vb*{u}' - \vb*{u}|+2|\vb*{u}|\cdot|\vb*{v}' - \vb*{v}|$.
\end{lem}
\begin{proof}
\textbf{(a)} 
Any scalar and vector product has the upper bound
$|\vb*{u}|\cdot|\vb*{v}|$. 
\begin{align*}
& |(\vb*{u}'\cdot \vb*{v}') - (\vb*{u}\cdot \vb*{v})|
=|(\vb*{u}'-\vb*{u})\cdot \vb*{v}' +\vb*{u}\cdot(\vb*{v}' -\vb*{v})| \leq \\
& \leq |(\vb*{u}'-\vb*{u})\cdot \vb*{v}'| +|\vb*{u}\cdot(\vb*{v}' -\vb*{v})| 
\leq |\vb*{u}'-\vb*{u}|\cdot|\vb*{v}'|+|\vb*{u}|\cdot|\vb*{v}' -\vb*{v}| = \\
& =|\vb*{u}' - \vb*{u}|+|\vb*{u}|\cdot|\vb*{v}' - \vb*{v}| \text{ due to } |\vb*{v}'|=1.
\end{align*}

\noindent
\textbf{(b)} 
Prove as (a) with the vector product instead of the scalar product.
\medskip

\noindent
\textbf{(c)} 
It follows by using $|\vb*{v}|=1$ and part (a):
\begin{align*}
& |(\vb*{u}'\cdot \vb*{v}')\vb*{v}' - (\vb*{u}\cdot\vb*{v})\vb*{v}|
=|(\vb*{u}'\cdot \vb*{v}'-\vb*{u}\cdot\vb*{v})\vb*{v}' + (\vb*{u}\cdot\vb*{v})(\vb*{v}'-\vb*{v})| \leq \\
& \leq  |\vb*{u}'\cdot \vb*{v}'-\vb*{u}\cdot\vb*{v}|\cdot |\vb*{v}'|
+|\vb*{u}\cdot\vb*{v}|\cdot |\vb*{v}'-\vb*{v}|\leq \\
& \leq |\vb*{u}'\cdot \vb*{v}'-\vb*{u}\cdot\vb*{v}|+|\vb*{u}|\cdot|\vb*{v}|\cdot|\vb*{v}'-\vb*{v}|
\leq |\vb*{u}' - \vb*{u}|+2|\vb*{u}|\cdot|\vb*{v}' - \vb*{v}|
\end{align*}
as required.
\end{proof}

Recall that $l_{N,A}$ denotes the minimum bond length between $N_i$ and $A_i$, and $L_{A,C}$ is the maximum distance between $A_i$ and $C_i$, while
$h$ is the minimum height in $\triangle N_i A_i C_i$ at $C_i$ for all residues in given backbones.

\begin{prop}[perturbations of a basis]
\label{prop:basis}
In the conditions of Theorem~\ref{thm:continuity},
if any atom is perturbed up to $\ep$, the basis vectors from Definition~\ref{dfn:bri} are perturbed as follows:
\smallskip

\noindent
\textbf{(a)} 
$|\vb*{u}'_i-\vb*{u}_i|\leq\dfrac{4\ep}{l_{N,A}}$;
\smallskip

\noindent
\textbf{(b)} 
$|\vb*{v}'_i-\vb*{v}_i|\leq\dfrac{8\ep}{h}(1+2\dfrac{L_{A,C}}{l_{N,A}})$; \smallskip

\noindent
\textbf{(c)} 
$|\vb*{w}'_i-\vb*{w}_i|\leq
4\ep K$, where $K=\dfrac{1}{l_{N,A}}+\dfrac{2}{h}\Big(1+2\dfrac{L_{A,C}}{l_{N,A}}\Big)$ for all $i=1,\dots,m$.
\end{prop}
\begin{proof}
\textbf{(a)} 
In Definition~\ref{dfn:bri} the vector $\vb*{u}_i=\dfrac{\ve{A_i N_i}}{|\ve{A_i N_i}|}$ satisfies the inequality $|\vb*{u}'_i-\vb*{u}_i|\leq\dfrac{4\ep}{l_{N,A}}$ by Lemma~\ref{lem:norm_vectors}.
\medskip

\noindent
\textbf{(b)} 
The second vector is $\vb*{v}_i=\dfrac{\vb*{h}_i}{|\vb*{h}_i|}$ for $\vb*{h}_i=\ve{A_i C_i}-b_i\ve{A_i N_i}$ and $b_i=\dfrac{\ve{A_i C_i}\cdot\ve{A_i N_i}}{|\ve{A_i N_i}|^2}$.
Set $\vb*{p}_i=\ve{A_i C_i}$, $\vb*{q}_i=\dfrac{\ve{A_i N_i}}{|\ve{A_i N_i}|}$, so $|\vb*{q}_i|=|\vb*{q}'_i|=1$, where any dash denotes a perturbation of a point or a vector.
Also, $|\vb*{p}_i|=|\ve{A_i C_i}|$ has the upper bound $L_{A,C}$. 
By Lemma~\ref{lem:products}(c): 
\begin{align*}
& |b'_i\ve{A'_i N'_i}-b_i\ve{A_i N_i}|=
|(\vb*{p}'_i\cdot \vb*{q}'_i)\vb*{q}'_i - (\vb*{p}_i\cdot \vb*{q}_i)\vb*{q}_i|
\leq \\
& |\vb*{p}'_i - \vb*{p}_i|+2|\vb*{p}_i|\cdot|\vb*{q}'_i - \vb*{q}_i|
\leq 2\ep+2L_{A,C}\dfrac{4\ep}{l_{N,A}},
\end{align*}
where we used $|\vb*{p_i}|\leq L_{A,C}$ and 
$|\vb*{q}'_i - \vb*{q}_i|\leq \dfrac{4\ep}{l_{N,A}}$ by Lemma~\ref{lem:norm_vectors}.
Then 
\begin{align*}
& |\vb*{h}'_i-\vb*{h}_i|=|\vb*{p}'_i-b'_i\ve{A'_i N'_i}-(\vb*{p}_i-b_i\ve{A_i N_i})| \leq \\
& \leq |\vb*{p}'_i-\vb*{p}_i|+|b'_i\ve{A'_i N'_i}-b_i\ve{A_i N_i}|\leq  2\ep+2\ep+\ep\dfrac{L_{A,C}}{l_{N,A}}=4\ep(1+2\dfrac{L_{A,C}}{l_{N,A}}).
\end{align*}
The vectors $\vb*{h}_i$, $\vb*{p}_i$, and $b_i\ve{A_i N_i}=(\vb*{p}_i\cdot\vb*{q}_i)\vb*{q}_i$ form a right-angled triangle with the hypotenuse $|\vb*{p}_i|$.
The length $|\vb*{h}_i|=|\ve{A_iC_i}|\sin\angle N_i A_i C_i$ is the height in $\triangle N_i A_i C_i$ at the atom $C_i$.
Using the given minimum height $h\leq |\vb*{h}_i|$, Lemma~\ref{lem:norm_vectors} for $\de=4\ep(1+2\dfrac{L_{A,C}}{l_{N,A}})$ implies that
$$|\vb*{v}'_i-\vb*{v}_i|
=\left| \dfrac{\vb*{h}'_i}{|\vb*{h}'_i|} - \dfrac{\vb*{h}_i}{|\vb*{h}_i|} \right|
\leq \dfrac{2\de}{h}\leq \dfrac{8\ep}{h}(1+2\dfrac{L_{A,C}}{l_{N,A}}).$$

\noindent
\textbf{(c)} 
The perturbation of  $\vb*{w}_i=\vb*{u}_i\times\vb*{v}_i$ is estimated by
Lemma~\ref{lem:products}(b):
\begin{align*}
& |\vb*{w}'_i-\vb*{w}_i|
=|(\vb*{u}'_i\times \vb*{v}'_i) - (\vb*{u}_i\times\vb*{v}_i)|\leq
|\vb*{u}'_i - \vb*{u}_i|+|\vb*{u}_i|\cdot|\vb*{v}'_i - \vb*{v}_i|\leq \\
& \leq\dfrac{4\ep}{l_{N,A}}+\dfrac{8\ep}{h}\Big(1+2\dfrac{L_{A,C}}{l_{N,A}}\Big) =4\ep K,
\text{ where }K=\dfrac{1}{l_{N,A}}+\dfrac{2}{h}\Big(1+2\dfrac{L_{A,C}}{l_{N,A}}\Big)
\end{align*}
as required.
\end{proof}
\smallskip

\begin{proof}[Proof of Theorem~\ref{thm:continuity}]
In the perturbed backbone $Q$, let $N'_i,A'_i,C'_i$ denote $\ep$-perturbations of atoms $N_,A_i,C_i$ from the backbone $S$ for $i=1,\dots,m$.
We prove that any coordinate of $\bri(S)$ changes by at most $\la\ep$ for the given Lipschitz constant $\la$.
The first coordinate $x(N_1)$ changes by at most $2\ep$ because $|x(N'_1)-x(N_1)|=\big|\, |\ve{A'_1N'_1}|-|\ve{A_1 N_1}|\,\big|\leq 2\ep$ by Lemma~\ref{lem:length_difference}.
For the coordinate $x(C_1)=\dfrac{\ve{A_1 C_1}\cdot\ve{A_1N_1}}{|\ve{A_1N_1}|}$, set $\vb*{u}=\ve{A_1 C_1}$ and $\vb*{v}=\dfrac{\ve{A_1N_1}}{|\ve{A_1N_1}|}$, so $|\vb*{v}|=1$.
We write the perturbed versions of all vectors with a dash.
\medskip

Then $|x(C'_1)-x(C_1)|=|\vb*{u}'\cdot\vb*{v}'-\vb*{u}\cdot\vb*{v}|\leq 
|\vb*{u}' - \vb*{u}|+|\vb*{u}|\cdot|\vb*{v}' - \vb*{v}|$ by Lemma~\ref{lem:products}(a).
Lemma~\ref{lem:vec_perturbation} implies that $|\vb*{u}' - \vb*{u}|\leq 2\ep$.
Lemma~\ref{lem:norm_vectors} for $u=\ve{A'_1N'_1}$ and $v=\ve{A_1N_1}$ implies that $|\vb*{v}' - \vb*{v}|\leq \dfrac{4\ep}{l_{N,A}}$, where $l_{N,A}$ is the minimum length of the bond between an $\al$-carbon $A_i$ and $N_i$ across all backbones.
Also, the Euclidean length $|\vb*{u}|=|\ve{A_1 C_1}|$ has the upper bound $L_{A,C}$ equal to the maximum length of the bond between $A_i$ and $C_i$ across all backbones.
Then $|x(C'_1)-x(C_1)|\leq 2\ep\Big(1+ 2\dfrac{L_{A,C}}{l_{N,A}}\Big)$.
\medskip

In the notations above, the last non-zero coordinate in the first row of $\bri(A)$ is
 $y(C_1)=|\ve{A_1 C_1} - x(C_1)\dfrac{\ve{A_1N_1}}{|\ve{A_1N_1}|}|=|\vb*{u}-x(C_1)\vb*{v}|$.
We estimate the perturbation first by Lemma~\ref{lem:length_difference}:  
\begin{align*}
& |y(C'_1)-y(C_1)|=|\,|\vb*{u}'-x(C'_1)\vb*{v}'| - |\vb*{u}-x(C_1)\vb*{v}| \,|
\leq \\
& \leq |\vb*{u}'-x(C'_1)\vb*{v}' - (\vb*{u}-x(C_1)\vb*{v})|
 \leq |\vb*{u}'-\vb*{u}|+|x(C'_1)\vb*{v}'-x(C_1)\vb*{v}| \leq \\
& \leq 2\ep+|(x(C'_1)-x(C_1))\vb*{v}'+x(C_1)(\vb*{v}'-\vb*{v})| \leq \\
& \leq 2\ep+|x(C'_1)-x(C_1)|+|x(C_1)|\cdot|\vb*{v}'-\vb*{v}| \leq \\
& \leq 2\ep+2\ep\Big(1+ 2\dfrac{L_{A,C}}{l_{N,A}}\Big)+|\ve{A_1 C_1}|\dfrac{4\ep}{l_{N,A}} 
\leq 4\ep \Big(1+2\dfrac{L_{A,C}}{l_{N,A}}\Big),
\end{align*}
where we substituted 
$|x(C'_1)-x(C_1)|\leq 2\ep(1+ 2\dfrac{L_{A,C}}{l_{N,A}})$ and
$|\vb*{v}'-\vb*{v}|\leq \dfrac{4\ep}{l_{N,A}}$. 
\smallskip

In any $i$-th row for $i=2,\dots,m$, we estimate 
by Proposition~\ref{prop:basis}(a):
\begin{align*}
& |x(N'_i)-x(N_i)|=|\ve{C'_{i-1}N'_i}\cdot \vb*{u}'_i - \ve{C_{i-1} N_i}\cdot \vb*{u}_i| \leq 
|\ve{C'_{i-1}N'_i} - \ve{C_{i-1} N_i}|+ \\
& + |\ve{C_{i-1} N_i}|\cdot |\vb*{u}'_i - \vb*{u}_i|
\leq 2\ep+L_{C,N}\dfrac{4\ep}{l_{N,A}}=2\ep(1+2\dfrac{L_{C,N}}{l_{N,A}})
\end{align*}
due to the upper bound $|\ve{C_{i-1} N_i}|\leq L_{C,N}$.
For the other coordinates $y,z$, similarly use Proposition~\ref{prop:basis}(b,c): 
\begin{align*}
& |y(N'_i)-y(N_i)|=|\ve{C'_{i-1}N'_i}\cdot \vb*{v}'_i - \ve{C_{i-1} N_i}\cdot \vb*{v}_i| \leq \\
& \leq |\ve{C'_{i-1}N'_i} - \ve{C_{i-1} N_i}|+|\ve{C_{i-1} N_i}|\cdot |\vb*{v}'_i - \vb*{v}_i|\leq \\
& \leq 2\ep+L_{C,N}\cdot \dfrac{8\ep}{h}(1+2\dfrac{L_{A,C}}{l_{N,A}})=2\ep\Big(1+4\dfrac{L_{C,N}}{h} (1+2\dfrac{L_{A,C}}{l_{N,A}}\Big). \\
& |z(N'_i)-z(N_i)|
=|\ve{C'_{i-1}N'_i}\cdot \vb*{w}'_i - \ve{C_{i-1} N_i}\cdot \vb*{w}_i|
\leq \\
& \leq  |\ve{C'_{i-1}N'_i} - \ve{C_{i-1} N_i}|+|\ve{C_{i-1} N_i}|\cdot |\vb*{w}'_i - \vb*{w}_i|\leq \\
& \leq 2\ep+L_{C,N}\cdot 4\ep K =2\ep(1+2L_{C,N}K),
\text{ for } K=\dfrac{1}{l_{N,A}}+\dfrac{2}{h}\Big(1+2\dfrac{L_{A,C}}{l_{N,A}}\Big).
\end{align*}

For the atoms $A_i,C_i$, we get similar upper bounds by replacing the factor $L_{C,N}$ with $L_{N,A}$, $L_{A,C}$, respectively.
Taking into account all upper bounds above, the overall upper bound for the $L_\infty$ metric on invariants is
$L_\infty(\bri(S),\bri(Q))\leq \la\ep$, where
$\la=2(1+2LK)$ for $L=\max\{L_{C,N},L_{N,A},L_{A,C}\}$ and
$K=\dfrac{1}{l_{N,A}}+\dfrac{2}{h}\Big(1+2\dfrac{L_{A,C}}{l_{N,A}}\Big)$ as required. 
\end{proof}

\begin{exa}[continuity in practice]
\label{exa:continuity}
For all 707K+ cleaned chains, the median upper bound for $\la$ is about 34.5 but the real values are smaller as in the example below.
Consider the backbone $S$ of the chain A (141 residues) from the standard hemoglobin 2hhb in the PDB.
\medskip

We perturb $S$ to $Q$ by adding to each coordinate $x,y,z$ of all atoms in $S$ some uniform noise up to various thresholds $\ep=0.01,0.02,\dots,0.1\angstrom$.
Fig.~\ref{fig:hemoglobins}~(top left) shows how the distance $L_\infty(\bri(S),\bri(Q))$ averaged over 20 perturbations depends on $\ep$
As expected by Theorem~\ref{thm:continuity}, the metric $L_\infty$ is perturbed linearly up to $\la\ep$, where $\la\approx 4$.
\end{exa}

Since the metric $L_\infty$ between invariants $\bri$ ($m\times 9$ matrices) can be computed in linear time $O(m)$, Theorem~\ref{thm:continuity} also completes condition (\ref{pro:map}f) in Problem~\ref{pro:map}.
Theorem~\ref{thm:inverse} will prove 
condition in \ref{pro:map}(d).

\begin{thm}[inverse continuity of $\bri$]
\label{thm:inverse}
For any $\de>0$ and backbones $S,Q\subset\R^3$ with $L_\infty(\bri(S),\bri(Q))<\de$, there is a rigid motion $f$ of $\R^3$ such that any atom of $S$ is $\mu\de$-close to the corresponding atom of $f(Q)$ for $\mu=\sqrt{3}\dfrac{(8LK)^{m-1}-1}{8LK-1}$.
Let $\widehat{\bri}(S)$ be $\bri(S)$ after multiplying the $i$-th row by $\dfrac{(8LK)^{i-1}-1}{8LK-1}$ for $i=2,\dots,m$.
Then $L_\infty(\widehat{\bri}(S),\widehat{\bri}(Q))<\de$ guarantees a rigid motion $f$ of $\R^3$ such that any atom of the backbone $S$ is $\sqrt{3}\de$-close to the corresponding atom of $f(Q)$.
\end{thm}
\begin{proof}[Proof of Theorem~\ref{thm:inverse}]
Choose the origin of $\R^3$ at the first alpha-carbon atom $A_1$ of the backbone $S$, the positive $x$-axis through the vector $\ve{A_1N_1}$, and the $y$-axis so that the triangle $N_1 A_1 C_1$ belongs to the upper half of the $xy$-plane.
Shift another backbone $Q$ so that its first alpha-carbon atom $A'_1$ coincides with the origin $A_1$. 
Rotate the image of $Q$ so that its first nitrogen atom $N_1$ is in the $x$-axis through the atoms $A_1,N_1$ of $S$ and the next carbon $C'_1$ of $Q$ is in the upper $xy$-plane.
\medskip

For the resulting motion $f$, we will prove that the atoms of $S$ are $\mu\de$-close to the corresponding atoms of the image of $Q$, which we still denote by $N'_i,A'_i,C'_i$ for simplicity.
Since the atom $N'_1$ is in the $x$-axis through $\ve{A_1N_1}$, the first basis vectors of length 1 coincide ($\vb*{u}'_1=\vb*{u}_1$) and hence also uniquely define the other basis vectors ($\vb*{v}'_1=\vb*{v}_1$, $\vb*{w}'_1=\vb*{w}_1$).
\medskip

Then $|x(N'_1)-x(N_1)|\leq\de$ implies that the atom $N'_1$ is $\de$-close to $N_1$ in the $x$-axis.
The atoms $C_1,C'_1$ are $\de\sqrt{2}$-close due to
$$|C'_1-C_1|=\sqrt{|x(C'_1)-x(C_1)|^2+|y(C'_1)-y(C_1)|^2}\leq\sqrt{\de^2+\de^2}=\de\sqrt{2}.$$ 
 
Since the first bases coincide, we consider the second residue:
\begin{align*}
& |N'_2-N_2|= \\
& =|x(N'_2)\vb*{u}_1 + y(N'_2)\vb*{v}_1 + z(N'_2)\vb*{w}_1 - 
x(N_2)\vb*{u}_1 - y(N_2)\vb*{v}_1 - z(N_2)\vb*{w}_1|
= \\
& =\sqrt{|x(N'_2)-x(N_2)|^2+|y(N'_2)-y(N_2)|^2+|z(N'_2)-z(N_2)|^2}\leq \\
& \leq\sqrt{\de^2+\de^2+\de^2}=\de\sqrt{3}.
\end{align*}

Similarly, we get the upper bound $\ep=\de\sqrt{3}$ for the deviations $|A'_2-A_2|$ and $|C'_2-C_2|$.
We will prove the following upper bound on deviations of atoms by induction on 
the number $m\geq 2$ of residues.   
$$\max\{|N'_m-N_m|,|A'_m-A_m|,|C'_m-C_m|\}\leq 
\sqrt{3}(1+8LK+\dots+8 (LK)^{m-2})\de, $$
$$\text{ where } 
L=\max\{L_{C,N},L_{N,A},L_{A,C}\},\quad 
K=\dfrac{1}{l_{N,A}}+\dfrac{2}{h}\Big(1+2\dfrac{L_{A,C}}{l_{N,A}}\Big).$$
The base $m=2$ was completed above.
The inductive assumption says that the upper bound $\ep=\sqrt{3}(1+8LK+\dots+(8 LK)^{i-2})\de$ holds for a single value of $i\geq 2$.
The inductive step below is for the next value $i+1$.
\medskip

Proposition~\ref{prop:basis} estimates deviations of vectors in the second basis:
$$|\vb*{u}'_2-\vb*{u}_2|\leq\dfrac{4\ep}{l_{N,A}},\quad
|\vb*{v}'_2-\vb*{v}_2|\leq\dfrac{8\ep}{h}\big(1+2\dfrac{L_{A,C}}{l_{N,A}}\big), \quad
|\vb*{w}'_2-\vb*{w}_2|\leq
4\ep K.$$ 
For nitrogens, we split the deviations in the $(i+1)$-st residue into the deviations proportional to the differences in coordinates and the deviations proportional to the differences in basis vectors as follows:
\begin{align*}
& |N'_{i+1}-N_{i+1}| =  |x(N'_{i+1})\vb*{u}'_i 
+y(N'_{i+1})\vb*{v}'_i + z(N'_{i+1})\vb*{w}'_i - \\
& - x(N_{i+1})\vb*{u}_i 
-y(N_{i+1})\vb*{v}_i - z(N_{i+1})\vb*{w}_i| = 
|(x(N'_{i+1})\vb*{u}'_i - x(N_{i+1})\vb*{u}_i) + \\
& +(y(N'_{i+1})\vb*{v}'_i - y(N_{i+1})\vb*{v}_i)
+(z(N'_{i+1})\vb*{w}'_i - z(N_{i+1})\vb*{w}_i)|= \\
& =\Big|
\big(x(N'_{i+1})- x(N_{i+1})\big)\vb*{u}'_i + x(N_{i+1})(\vb*{u}'_i-\vb*{u}_i)
+\big(y(N'_{i+1})- y(N_{i+1})\big)\vb*{v}'_i +\\ 
& + y(N_{i+1})(\vb*{v}'_i-\vb*{v}_i)
+\big(z(N'_{i+1})- z(N_{i+1})\big)\vb*{w}'_i + z(N_{i+1})(\vb*{w}'_i-\vb*{w}_i)\Big|\leq \\
& \leq \Big| \big(x(N'_{i+1})- x(N_{i+1})\big)\vb*{u}'_i+ \big(y(N'_{i+1})- y(N_{i+1})\big)\vb*{v}'_i+ \\
& + \big(z(N'_{i+1})- z(N_{i+1})\big)\vb*{w}'_i \Big| +\Big| x(N_{i+1})(\vb*{u}'_i-\vb*{u}_i)\Big| + \\
& + \Big| y(N_{i+1})(\vb*{v}'_i-\vb*{v}_i)\Big|
+ \Big| z(N_{i+1})(\vb*{w}'_i-\vb*{w}_i)\Big|.
\end{align*}

In the last upper bound, the first big modulus is the Euclidean length of a vector written in the orthonormal basis $\vb*{u}'_i, \vb*{v}'_i, \vb*{w}'_i$.
Since the coordinates of this vector have absolute values at most $\de$, this length has the upper bound $\de\sqrt{3}$.
In the second row of the matrix $\bri$, we estimate each term by replacing absolute values of coordinates with the maximum bond lengths and by using
$|x(N_{i+1})|\leq L_{C,N}$ and Proposition~\ref{prop:basis} as follows: 
\begin{align*}
& |x(N_{i+1})|\cdot |\vb*{u}'_i-\vb*{u}_i|\leq 
 L_{C,N}\dfrac{4\ep}{l_{N,A}}, \\
& |y(N_{i+1})|\cdot |\vb*{v}'_i-\vb*{v}_i|\leq 
 L_{C,N}\dfrac{8\ep}{h}\Big(1+2\dfrac{L_{A,C}}{l_{N,A}}\Big), \\
& |z(N_{i+1})|\cdot |\vb*{w}'_i-\vb*{w}_i|\leq 
 L_{C,N}\cdot 4\ep K, \text{ where }
 K=\dfrac{1}{l_{N,A}}+\dfrac{2}{h}\Big(1+2\dfrac{L_{A,C}}{l_{N,A}}\Big).
\end{align*}

Taking the sum of these estimates,  the final deviation 
is
\begin{align*}
& |N'_{i+1}-N_{i+1}|\leq \\
& \sqrt{3}\de+ 4\ep L_{C,N}\Big(\dfrac{1}{l_{N,A}}
 +\dfrac{2}{h}\big(1+2\dfrac{L_{A,C}}{l_{N,A}}\big)
 +\dfrac{1}{l_{N,A}}+\dfrac{2}{h}\big(1+2\dfrac{L_{A,C}}{l_{N,A}}\big)
 \Big)= \\
& =\sqrt{3}\de+8L_{C,N}\ep K\leq \sqrt{3}(1+ 8LK(1+\dots+(8LK)^{i-2})\de
 = \\
& =\sqrt{3}(1+\dots+(8LK)^{i-1})\de.
\end{align*}

For the atoms $A_{i+1},C_{i+1}$ in the $(i+1)$-st residue, we get the same bound by replacing $L_{C,N}$ with $L_{N,A},L_{A,C}\leq L$.
The bound for $i=m$ is 
$\sqrt{3}(1+\dots+(8LK)^{m-2}))\de=\sqrt{3}\dfrac{(8LK)^{m-1}-1}{8LK-1}\de$.
\medskip

Now consider the modified invariant $\widehat{\bri}(S)$ obtained by multiplying the $i$-th row of $\bri(S)$ by $\dfrac{(8LK)^{i-1}-1}{8LK-1}$ for $i=2,\dots,m$.
Then the $\de$-closeness of the corresponding invariant components in the metric $L_\infty$ means smaller deviations $|x(N'_i)-x(N_i)|\leq\de\dfrac{8LK-1}{(8LK)^{i-1}-1}$, similarly for other components.
This extra multiplicative factor gives the upper bound 
$|N'_{i+1}-N_{i+1}|\leq\sqrt{3}\de$, similarly for all other atoms.
\end{proof}

A Lipschitz constant $\mu$ plays no significant role because any metric on invariant values can be divided by $\mu$, which makes this constant 1.  
The second part of Theorem~\ref{thm:inverse} offers a smarter adjustment of $\bri(S)$ to the modified invariant $\widehat{\bri}(S)$ depending on a row index of $\bri(S)$ to guarantee the smaller Lipschitz constant $\sqrt{3}$.

\section{Average invariant, diagrams, barcodes}
\label{sec:barcode}

This section simplifies the complete invariant $\bri$ to its average vector in $\R^9$ and introduces the diagram and barcode that visually represent $\bri$. 

\begin{dfn}[average invariant $\brain$, standard deviation of invariants, diagram $\bid$, and barcode $\bib$]
\label{dfn:diagrams}
For any protein backbone $S$ of $m$ residues, the \emph{backbone rigid average invariant} $\brain(S)\in\R^9$ is the vector of nine column averages in $\bri(S)$ excluding the first row. The standard deviation can be computed in a similar way.
The \emph{backbone invariant diagram} $\bid(S)$ consists of nine polygonal curves going through the points $(i,c(i))$, $i=2,\dots,m$, where $c$ is one of the coordinates (columns) of $\bri(S)$, see Fig.~\ref{fig:hemoglobins}~(middle).
For each atom type such as $N$, the coordinates $(x(N_i),y(N_i),z(N_i))$ are linearly converted into the RGB color value for $i=1,\dots,m$.
The resulting three color bars for the ordered atoms $N,A,C$ form the \emph{backbone invariant barcode} $\bib(S)$, see Fig.~\ref{fig:hemoglobins}~(bottom).
\end{dfn}

\begin{exa}[hemoglobins]
\label{exa:hemoglobins}
The PDB contains thousands of hemoglobin structures.
We consider here the structure 2hhb as a standard, and compare it with oxygenated 1hho, which contains an extra oxygen whose transport is facilitated by hemoglobin.
In both cases, we considered the main chains (entity 1, model 1, chain A) of 141 residues. 
Table \ref{tab:haemoglobin_invariants} showed the TRIN and BRI invariants for the first 3 residues of 2hhb and 1hho.
\smallskip

\begin{figure}[ht!]
\includegraphics[height=18.5mm]{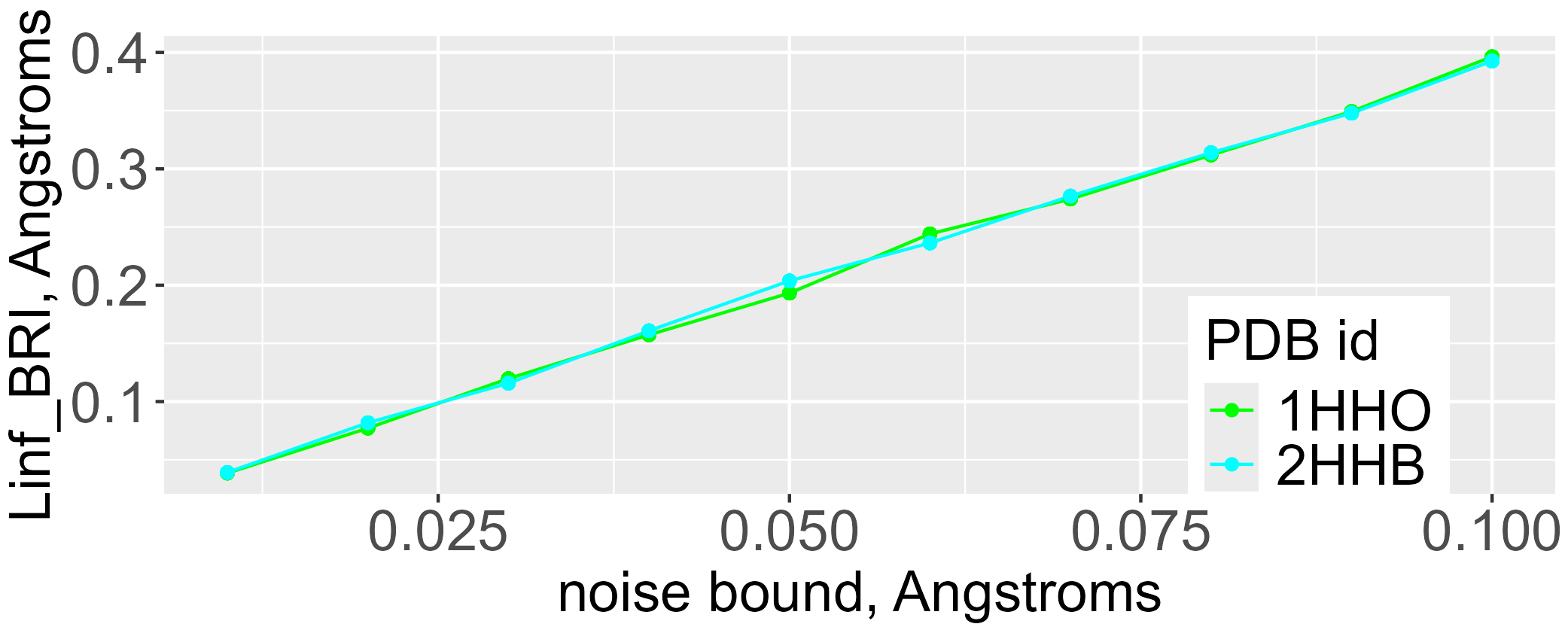}
\includegraphics[height=18.5mm]{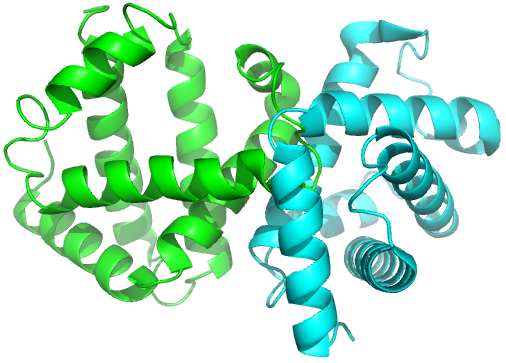}
\includegraphics[height=18.5mm]{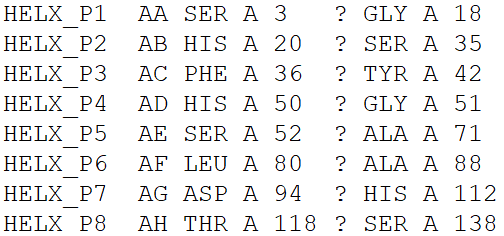}
\medskip

\includegraphics[height=36mm]{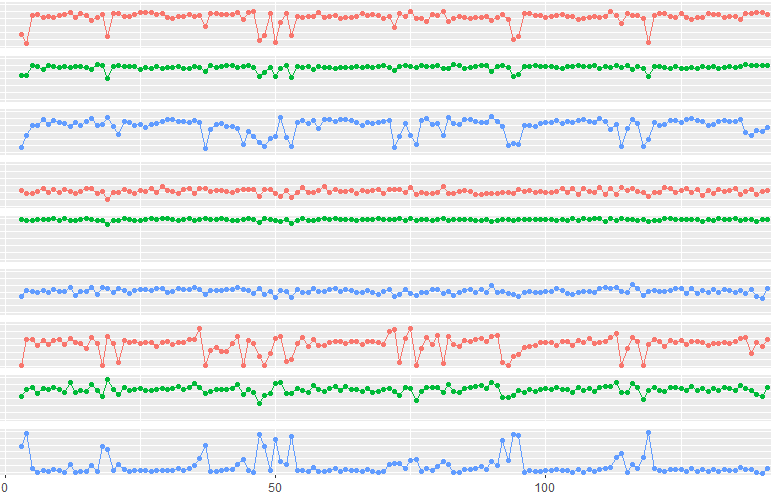}
\includegraphics[height=36mm]{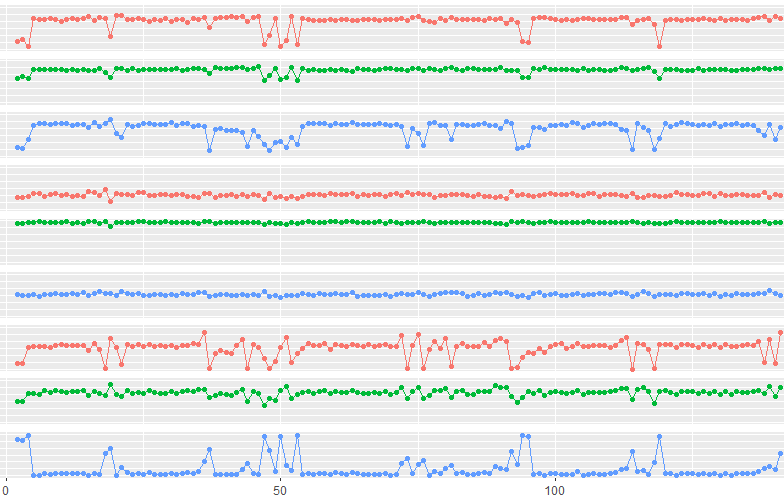}
\medskip

\includegraphics[height=13.5mm]{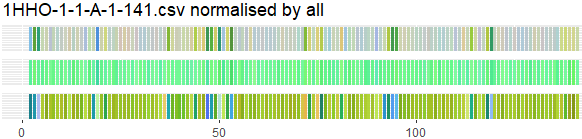}
\includegraphics[height=13.5mm]{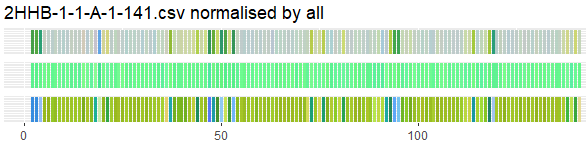}

\caption{
\textbf{Row 1}: the Lipschitz continuity of $\bri$ from Theorem~\ref{thm:continuity} is illustrated on the left by perturbing hemoglobins in
Example~\ref{exa:continuity}, whose main chains A of 141 residues are shown in the middle (oxygenated 1hho in green, standard 2hhb in cyan) and eight $\al$-helices found by \cite{kabsch1983dictionary} 
on the right.
\textbf{Row 2}: the Backbone Invariant Diagram ($\bid$) of the hemoglobins 1hho vs 2hhb in the PDB, see Definition~\ref{dfn:diagrams}.
\textbf{Row 3}: the Backbone Invariant Barcode ($\bib$), see Example~\ref{exa:hemoglobins}. }
\label{fig:hemoglobins}
\end{figure}

The top left image in Fig.~\ref{fig:hemoglobins}~(top) shows that the Lipschitz constant from Theorem~\ref{thm:continuity} is $\la\approx 4$ for both hemoglobins.
Fig.~\ref{fig:hemoglobins}~(middle) illustrates the complexity of identifying similar proteins that can be given with very distant coordinates.
The similarity under rigid motion becomes clearer by comparing their diagrams and barcodes in Fig.~\ref{fig:hemoglobins}~(rows 2, 3).
\smallskip

More importantly, a rigidly repeated pattern such as $\al$-helix or $\be$-strand has constant invariants over several residue indices, which are easily detectable in $\bid$ and visible in $\bib$ as intervals of uniform color.
The PDB uses the baseline algorithm DSSP (Define Secondary Structure of Proteins) \cite{kabsch1983dictionary}, which depends on several manual parameters and sometimes outputs $\al$-helices of only two residues.
For instance, the PDB files 1hho and 2hhb in Fig.~\ref{fig:hemoglobins}~(right) include HELX\_P4  consisting of only residues 50 and 51, and HELX\_P5  of length 20 over residue indices $i=52,\dots,71$.
Fig.~\ref{fig:hemoglobins} shows that a `constant' interval of little noise appears only for $i=54,\dots,70$.
Hence new invariants allow a more objective detection of secondary structures,  which will be explored in future work.
\end{exa}

\begin{figure}[ht!]
\includegraphics[width=0.325\textwidth]{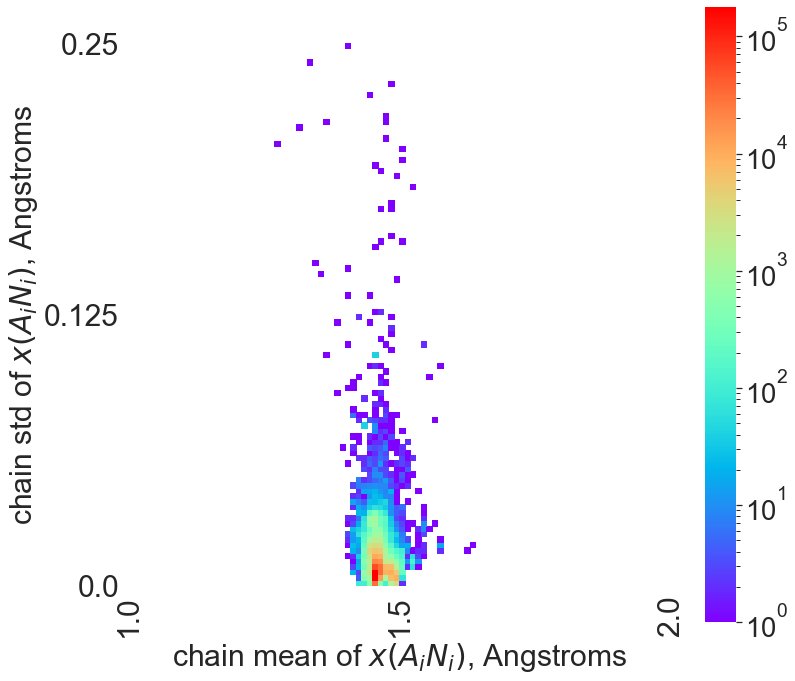}
\includegraphics[width=0.325\textwidth]{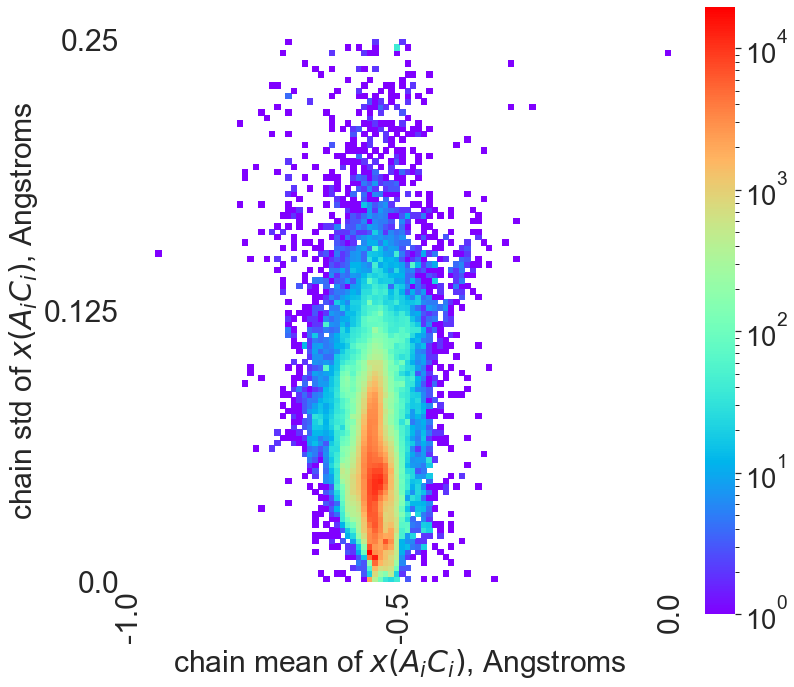}
\includegraphics[width=0.325\textwidth]{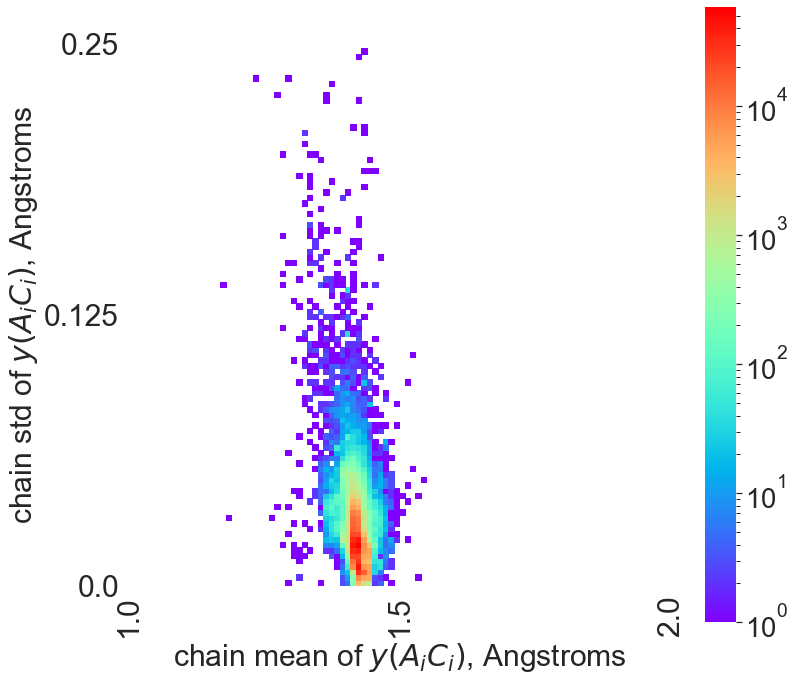}
\medskip

\includegraphics[width=0.325\textwidth]{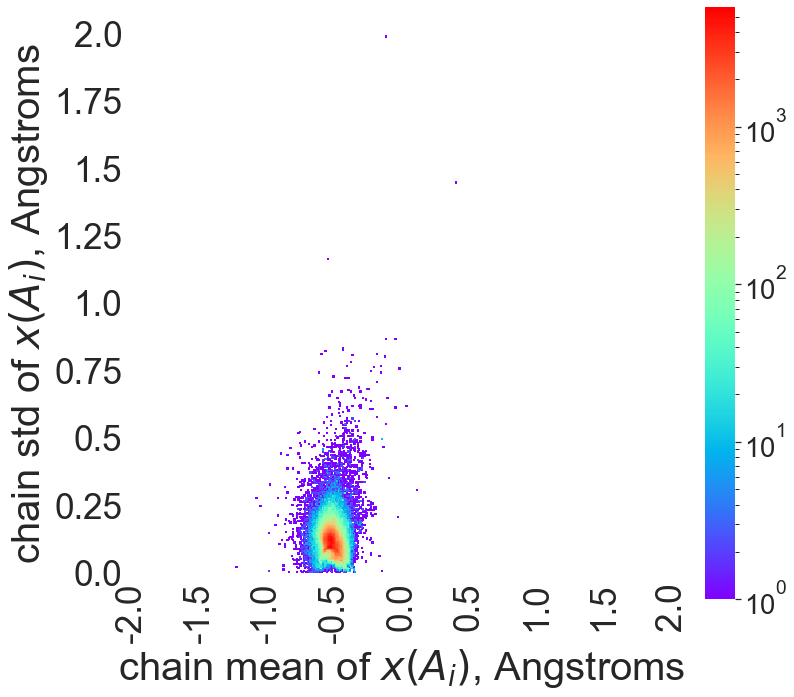}
\includegraphics[width=0.325\textwidth]{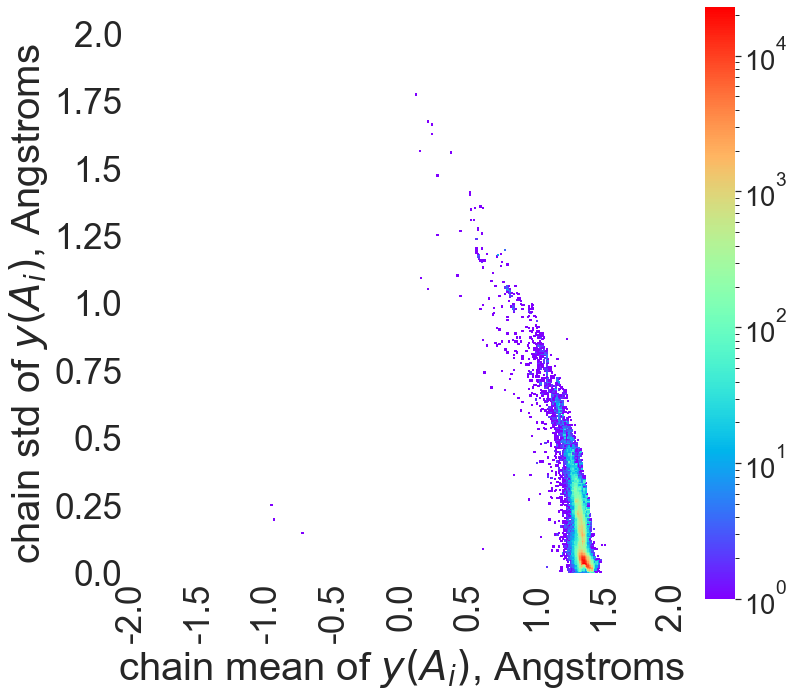}
\includegraphics[width=0.325\textwidth]{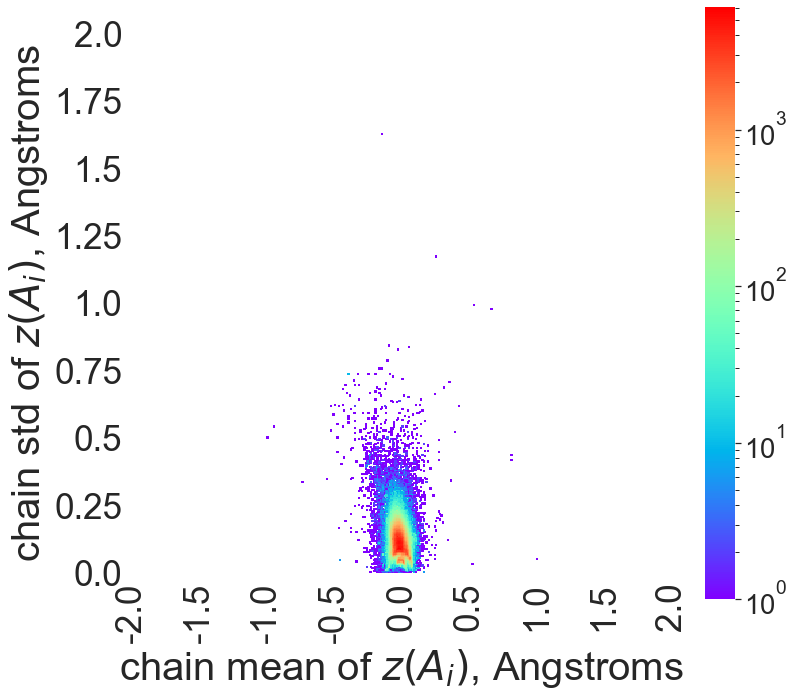}
\medskip

\includegraphics[width=0.325\textwidth]{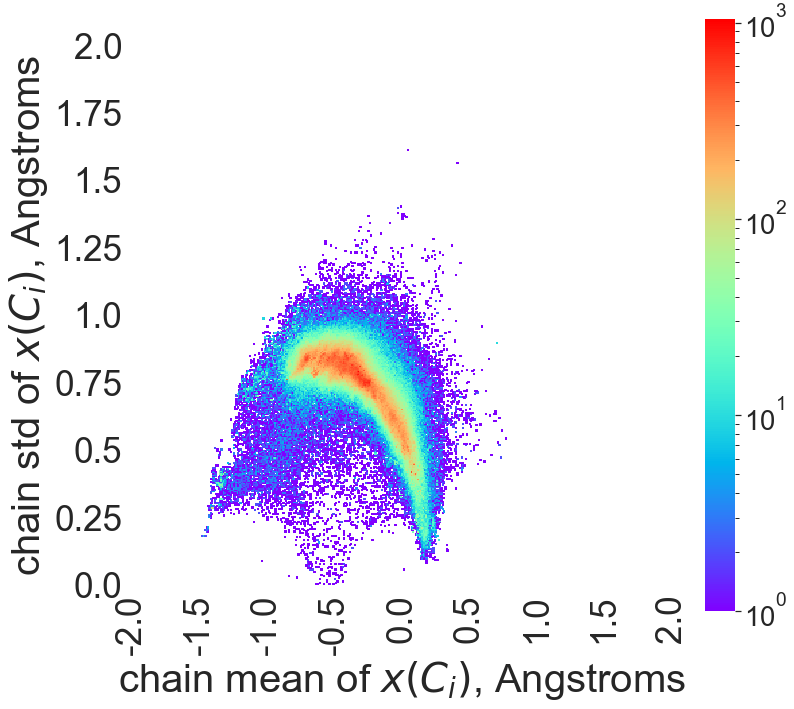}
\includegraphics[width=0.325\textwidth]{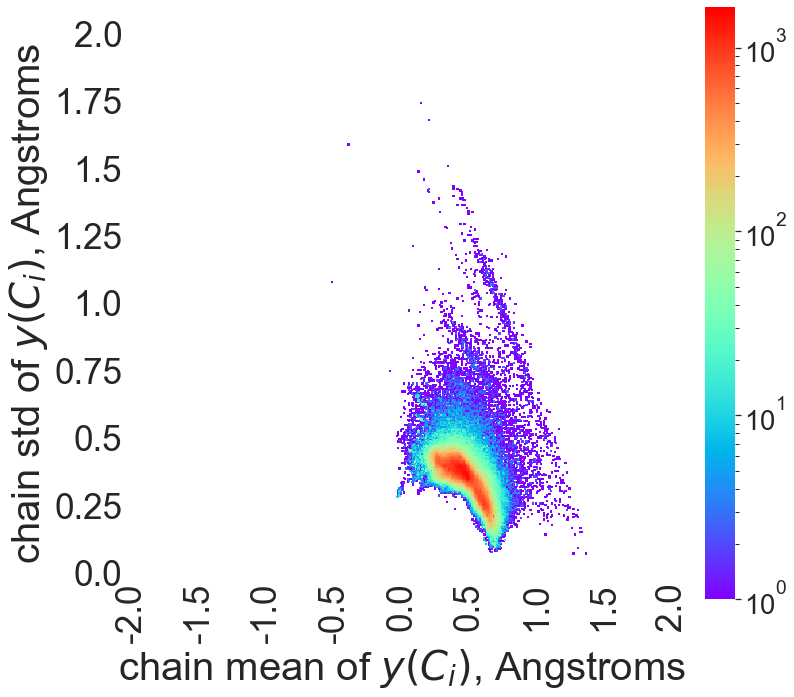}
\includegraphics[width=0.325\textwidth]{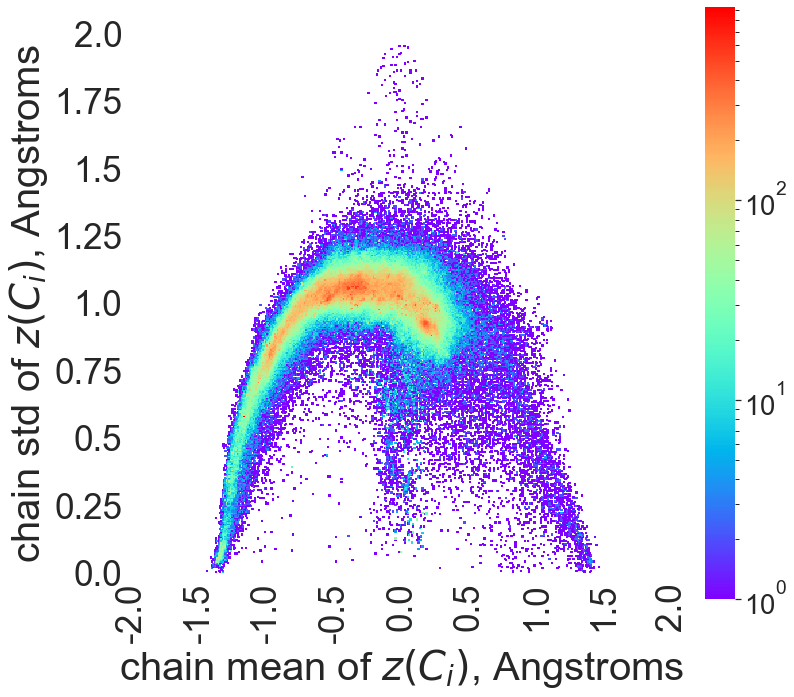}
\medskip

\includegraphics[width=0.325\textwidth]{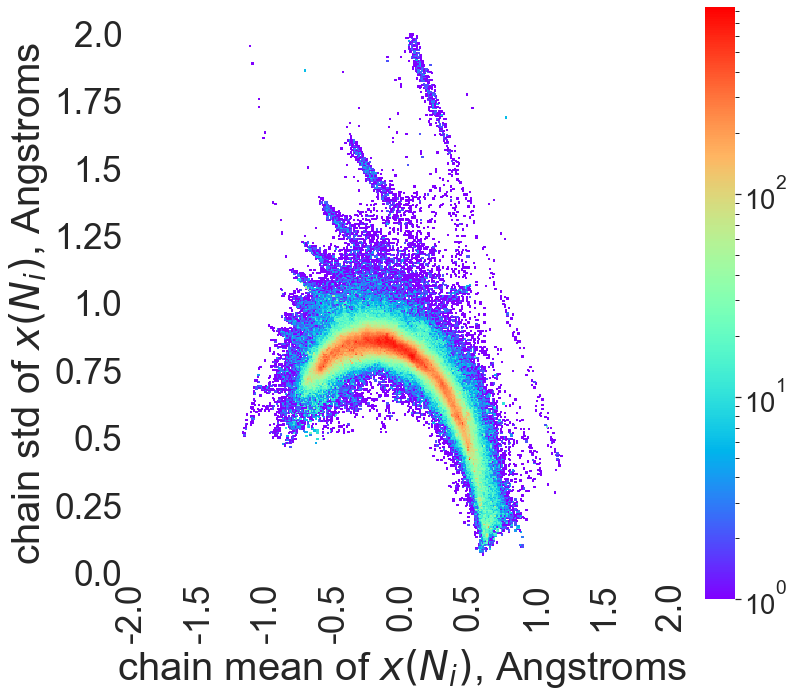}
\includegraphics[width=0.325\textwidth]{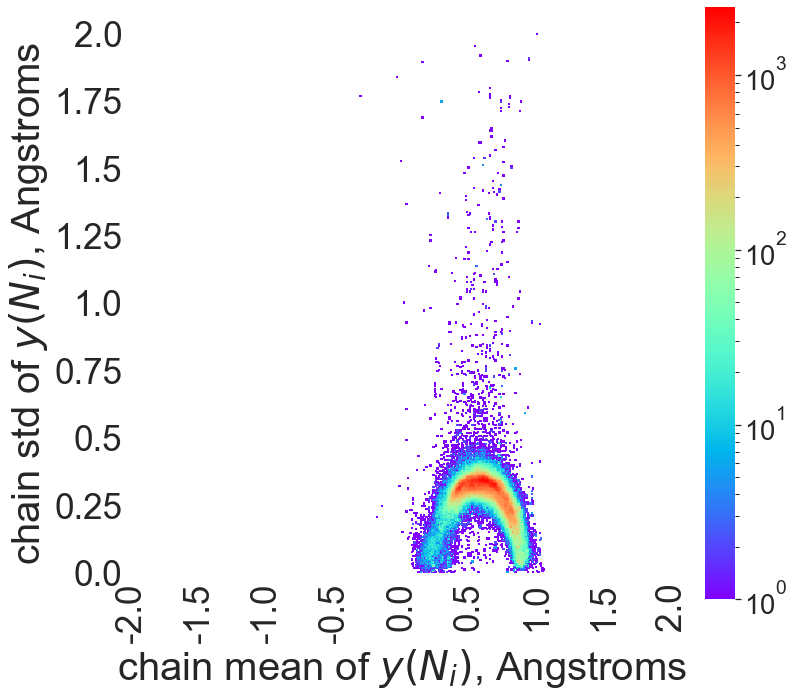}
\includegraphics[width=0.325\textwidth]{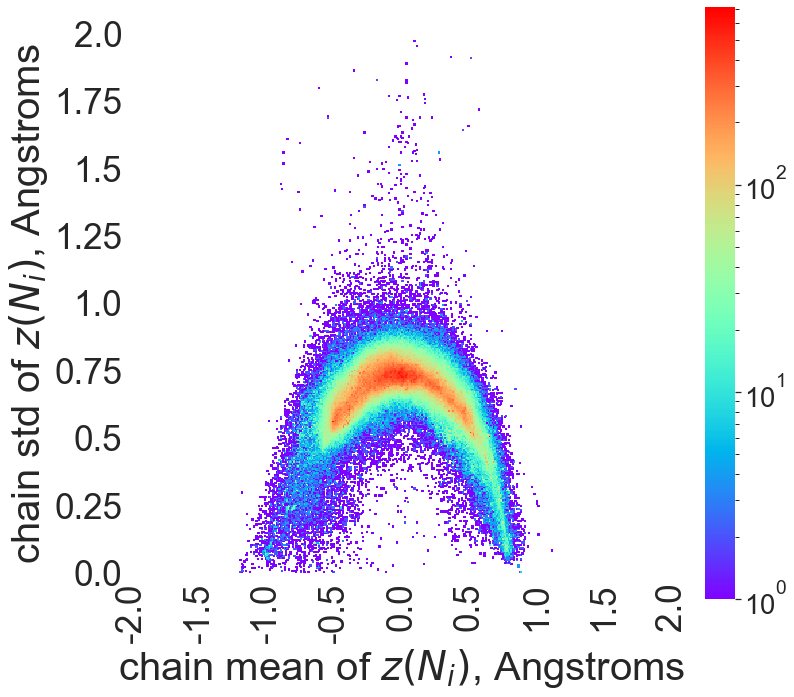}

\caption{
Heatmaps of average and standard deviations of the invariants $\trin$ and $\bri$ 
across all 707K+ chains obtained by Protocol~\ref{prot:cleanPDB}.
The color indicates (on the logarithmic scale) the number of chains whose pairs $(x,y)$ of the average $x$ and the standard deviation $y$ are discretized to each pixel.
}
\label{fig:PDB707K_BRI_heatmaps_by_chain_log}
\end{figure}

While the complete $\bri(S)$ can be used to compare backbones of the same length, the average invariant $\brain(S)\in\R^9$ and the standard deviation invariant can help to visualize all backbones of different lengths on the same heatmap.
In each image of Fig.~\ref{fig:PDB707K_BRI_heatmaps_by_chain_log}, any protein backbone is represented by a single point $(x,y)$ whose coordinates are the two simplest statistics (average and standard deviation) of a fixed invariant across all residues in a fixed chain.  
The top images in Fig.~\ref{fig:PDB707K_BRI_heatmaps_by_chain_log} show that the deviations of all three invariants from Definition~\ref{dfn:trin} can be as large as $0.25\angstrom$.
So the shapes of residue triangles can substantially vary even for a fixed backbone, while AlphaFold2 \cite{jumper2021highly} assumes that they all have identical shapes.  

\section{Duplicates with identical coordinates}
\label{sec:duplicates}

The linear time of the complete invariant $\bri(S)$ has enabled all-vs-all comparisons for all tertiary structures in the PDB, which was additionally cleaned by Protocol~\ref{prot:cleanPDB}.
To speed up comparisons, Lemma~\ref{lem:brain} proves that the metric  $L_\infty(\bri(S),\bri(Q))$ between complete invariants is not smaller than 
the much faster distance $L_\infty(\brain(S),\brain(Q))$ between the averaged invariants (vectors of 9 coordinates) from Definition~\ref{dfn:diagrams}.

\begin{lem}[relation between metrics on the invariants $\bri$ and $\brain$]
\label{lem:brain}
Any protein backbones $S,Q$ of the same number of residues
satisfy the inequality $L_\infty(\brain(S),\brain(Q))\leq L_\infty(\bri(S),\bri(Q))$.
\end{lem}
\begin{proof}[Proof of Lemma~\ref{lem:brain}]
If protein backbones $S,Q$ have $m$ residues and $\de=L_\infty(\bri(S),\bri(Q))$, then any corresponding elements of the $m\times 9$ matrices $\bri(S),\bri(Q)$ differ by at most $\de$.
For any $j=1,\dots,9$, their averages of the $j$-th columns differ by at most $\de$ because
\begin{align*}
& \left|\dfrac{1}{m}\sum\limits_{i=1}^m \bri_{ij}(S)-\dfrac{1}{m}\sum\limits_{i=1}^m \bri_{ij}(Q)\right| \leq \\
& \dfrac{1}{m}\sum\limits_{i=1}^m |\bri_{ij}(S)-\bri_{ij}(Q)| 
\leq \dfrac{1}{m}\sum\limits_{i=1}^m\de=\de.
\end{align*}
Hence $L_\infty(\brain(S),\brain(Q))\leq\de$ as required.
\end{proof}

The complete invariants and their statistical summaries (averages and deviations) were computed in 3 hours 18 min 21 sec.
After comparing all (888+ million) pairs of same-length backbones within 1 hour, we found 13907 pairs $S,Q$ with the \emph{exact zero-distance} $L_\infty(\bri(S),\bri(Q))=0$ between complete invariants meaning that all these backbones $S,Q$ are related by rigid motion, but they may not be geometrically identical.
\medskip

However, 9366 of these pairs turned out to have $x,y,z$ coordinates of all main atoms \emph{identical to the last digit} despite many of them (763) coming from \emph{different PDB entries}. 
Table~\ref{tab:duplicates} lists nine pairs whose geometrically identical chains unexpectedly differ in the sequences of amino acids.
\medskip

In a similar case \cite{widdowson2022resolving}, when five pairs of unexpected duplicates were found in the Cambridge Structural Database (CSD).
Their integrity office agreed that a single atomic replacement should perturb geometry at least slightly, so all coordinates cannot remain the same.
Five journals started investigations into the data integrity of the relevant publications \cite{chawla2024crystallography}.  
\medskip

We e-mailed all authors of the experimental structures listed in Table~\ref{tab:duplicates} whose contacts we found.
Two authors replied with details and confirmed that their PDB entries should be corrected, see appendix~\ref{sec:appendix}. 
\medskip

\begin{table}
\caption{Chains with identical backbones but different sequences.} 
\label{tab:duplicates}
\centering
\begin{tabular}{lllll}
PDB id1 & method and & PDB id2  & 
all atoms have  & different\\
\& chain & resolutions, $\angstrom$ & \& chain & 
identical $x,y,z$ & residues \\
\hline
1a0t-B  & X-ray, 2.4, 2.4
 & 1oh2-B 
 & all $3\times 413$ & 9 \\
1ce7-A & X-ray, 2.7, 2.7
 & 2mll-A 
 & all $3\times 241$ & 1, GLY$\neq$HIS  \\
1ruj-A & X-ray, 3, 3
 & 4rhv-A 
 & all $3\times 237$ & 1, GLY$\neq$SER \\
1gli-B/D & X-ray, 2.5, 1.7
 & 3hhb-B/D 
 & all $3\times 146$ & 1, MET$\neq$VAL \\
2hqe-A & X-ray, 2, 2
 & 2o4x-A 
 & all $3\times 217$ & 1, GLN$\neq$GLU \\
5adx-T &  EM, 4, 8.2
 & 5afu-Z 
 & all $3\times 165$ & 1, ILE$\neq$VAL \\
5lj3-O & EM, 3.8, 10
 & 5lj5-P 
 & all $3\times 252$ & 1, ALA$\neq$VAL \\
8fdz-A & X-ray, 2.5, 2.2
 & 8fe0-A 
 & all $3\times 200$ & 1, THR$\neq$SER \\
\end{tabular}
\end{table}


The duplicates from Table~\ref{tab:duplicates} were shown to the PDB validation team, who did not know about the found coincidences (in coordinates) and differences (in amino acids) because the PDB validation is currently done for an individual protein (checking atom clashes, outliers etc).
\medskip

The recently published method \cite{guzenko2020real} didn't report any duplicates. 
Right now anyone can download the PDB files from Table~\ref{tab:duplicates} and see all coincidences of $x,y,z$ coordinates with their own eyes without any computations.
Here are the links to the identical files in the first row of Table~\ref{tab:duplicates}, where the 4-letter PDB id can be replaced with any other id:
https://files.rcsb.org/download/1A0T.cif and \\
https://files.rcsb.org/download/1OH2.cif.
\medskip

\begin{figure}[h!]
\centering
\includegraphics[width=0.49\textwidth]{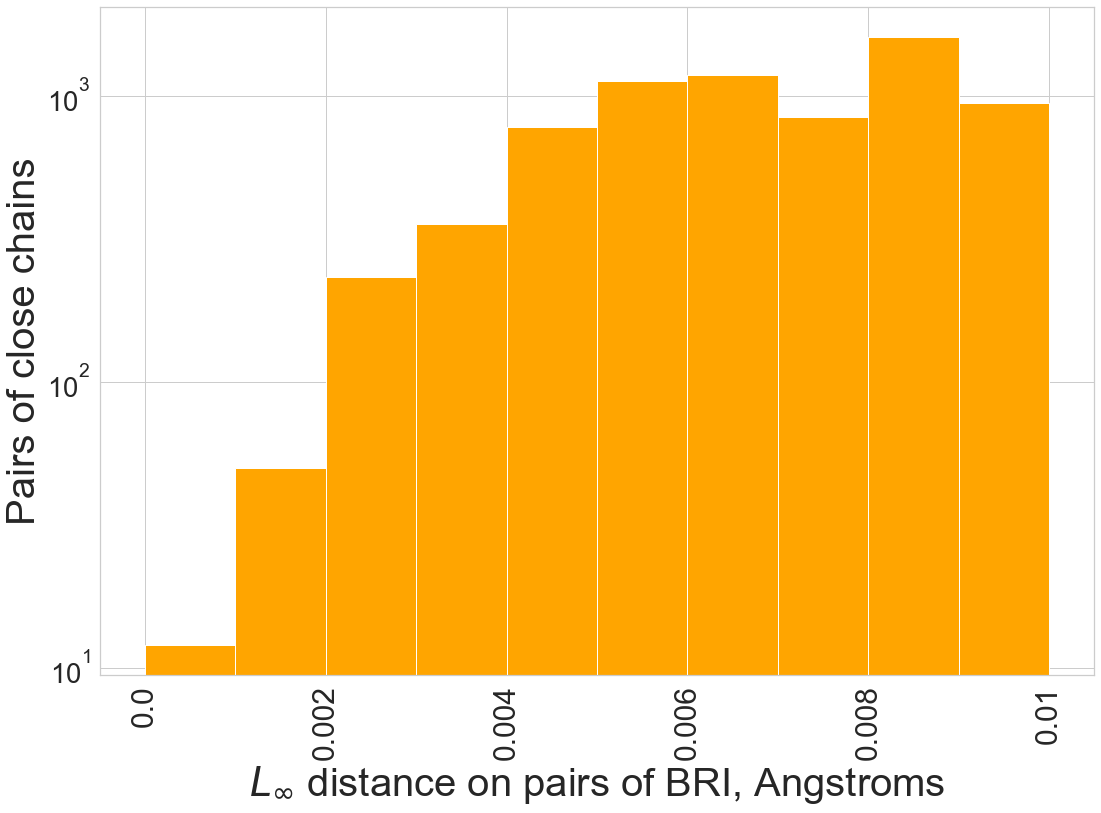}
\includegraphics[width=0.49\textwidth]{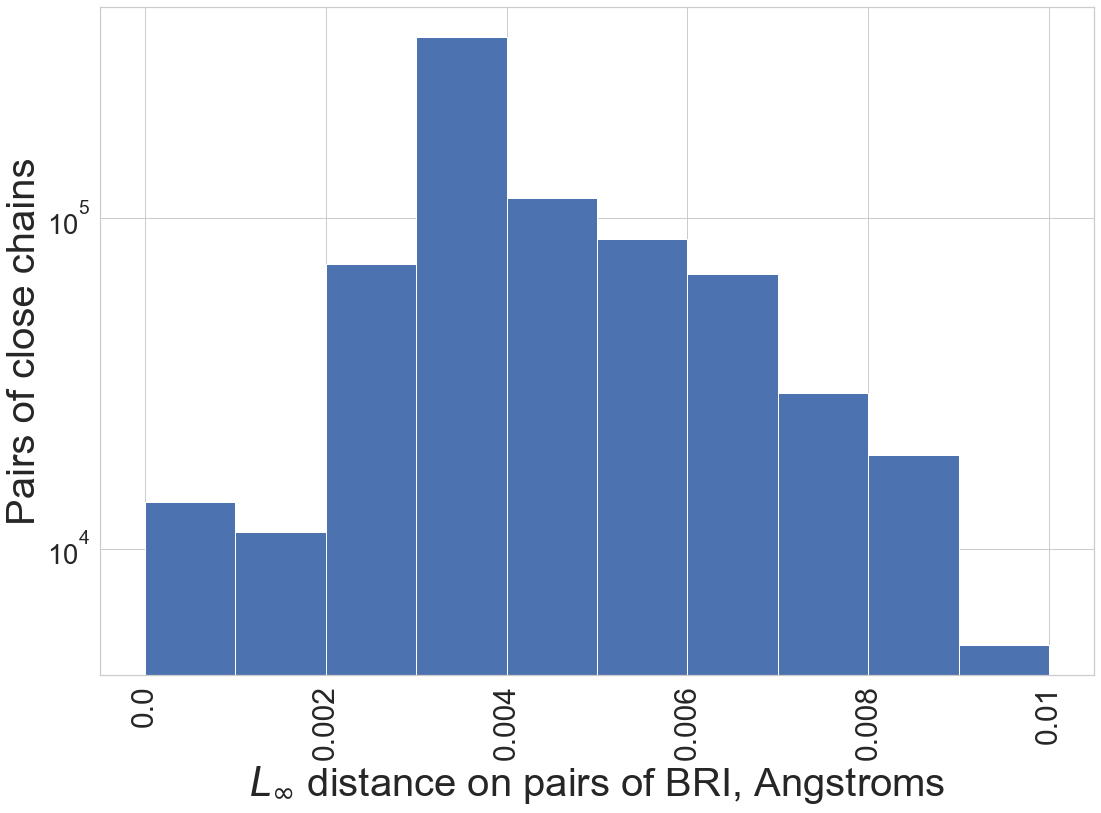} 

\includegraphics[width=0.49\textwidth]{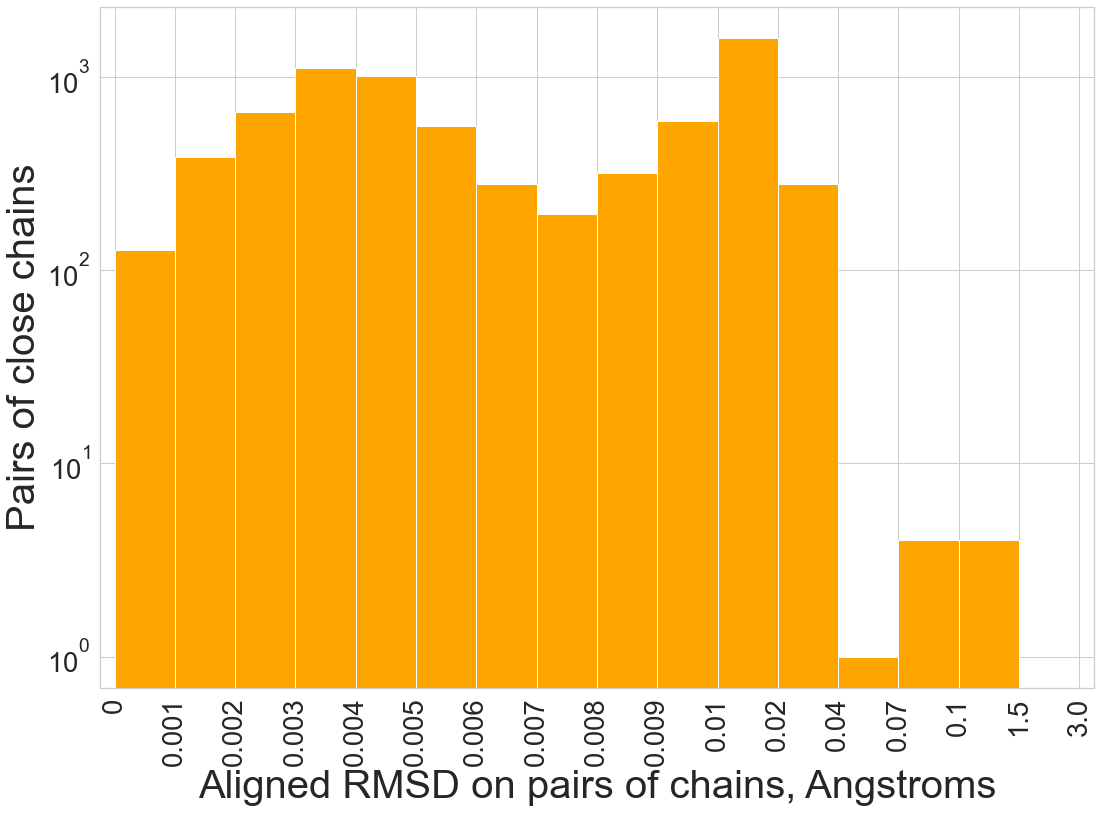}
\includegraphics[width=0.49\textwidth]{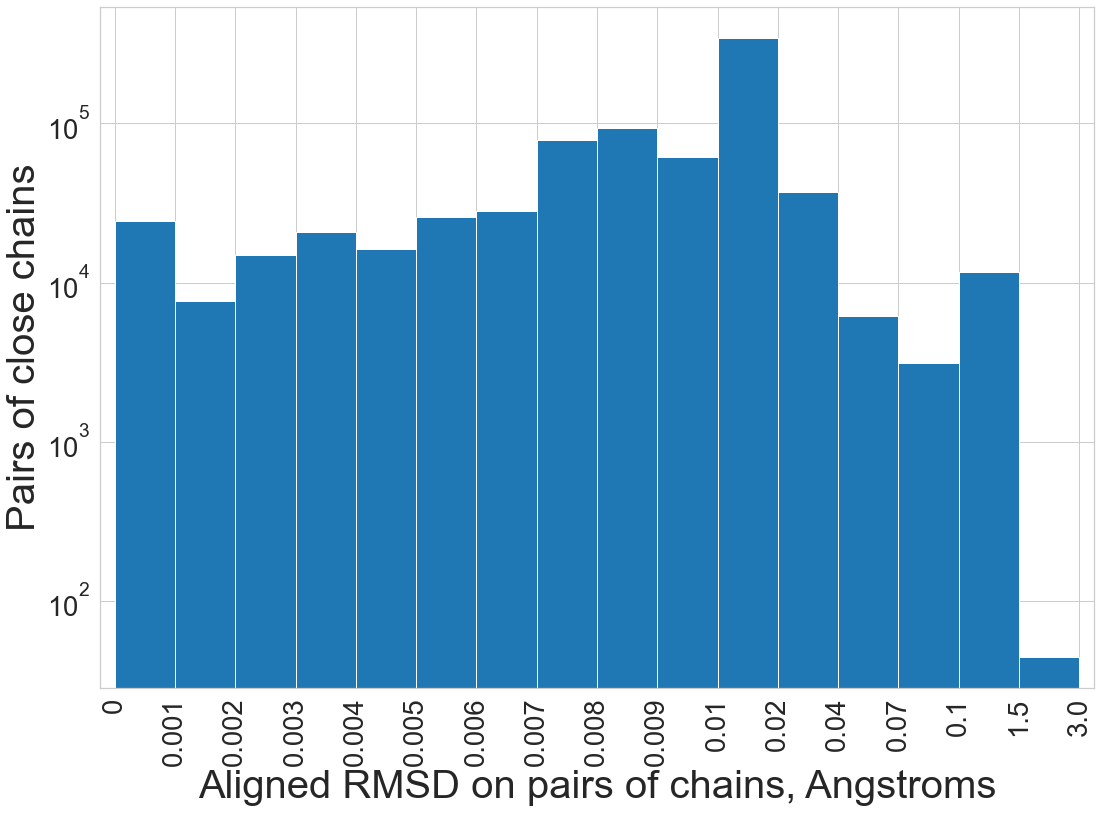}
\caption{
Histograms of near-duplicate chains (of the same length) on the log scale. 
\textbf{Top row}: 783075 pairs with $L_\infty\leq 0.01\angstrom$ on $\bri$s  including 13907 pairs of exact duplicates with $L_\infty=0$.
\textbf{Bottom row}: the same pairs with traditional RMSD.
\textbf{Left column}:  7151 pairs with different sequences of amino acids.
\textbf{Right column}: 775852 pairs with identical sequences.
}
\label{fig:near-duplicates}
\end{figure}

The histogram in Fig.~\ref{fig:near-duplicates} reveals the scale of near-duplicates among 707K+ cleaned chains up to small distances $L_\infty\leq 0.01\angstrom$ on the horizontal axis.
Each of 10 vertical bins over an interval of length $0.001\angstrom$ indicates the number of pairs (on the logarithmic scale) of backbones $S,Q$ whose distance $L_\infty(\bri(S),\bri(Q))$ is within this interval.
Since all atomic coordinates in the PDB have 3 decimal places, all distances were rounded to $0.001\angstrom$.
The bound of $0.01\angstrom$ is considered noise because the smallest inter-atomic distance is about 100 times larger at $1\angstrom=10^{-10}$ m.
\medskip

The physical meaning of distances follows from the bi-continuity conditions (c,d) in Problem~\ref{pro:map}.
If every atom of a backbone $S$ is shifted up to Euclidean distance $\ep$, then $\bri(S)$ changes up to $\la\ep$ in $L_\infty$.
The Lipschitz constant $\la$ was expressed in Theorem~\ref{thm:continuity} and estimated as $\la\approx 4$ for the hemoglobin chains in Example~\ref{exa:hemoglobins}.
So any small perturbation of atoms yields a small value of $L_\infty$ in Angstroms. 
The inverse Lipschitz continuity in (\ref{pro:map}d) implies that a small distance $L_\infty(\bri(S),\bri(Q))=\de$ guarantees that all atoms of $S,Q$ can be matched (after a suitable rigid motion) up to Euclidean distance $\mu\de$, see Theorem~\ref{thm:inverse}.
For all 775K+ pairs in Fig.~\ref{fig:near-duplicates}, the median of the maximum atomwise deviation (of optimally aligned chains) divided by $L_\infty(\bri(S),\bri(Q))$ is about $0.4$. 
So the closeness of $\bri$s practically guarantees the closeness by $\RMSD$.
\medskip

One potential explanation of identical coordinates is the molecular replacement method \cite{rossmann1990molecular}, which uses an existing protein structure, often a previous PDB deposit or part thereof, to solve a new structure. 
If the newly calculated electron density map does not allow for further refinement then the coordinates may (reasonably) remain unchanged.
\medskip
 
The same coincidences can happen with lower-quality cryo-EM maps in which an existing PDB structure may be placed but where the resolution may not allow for further refinement of atomic coordinates \cite{murshudov2011refmac5,hekkelman2024pdbredo}.
\medskip

We have checked that the found duplicate backbones also have identical distance matrices on $3m$ ordered atoms, which were slower to compute in time $O(m^2)$ over two days on a similar machine.
The widely used DALI server \cite{holm2024dali} also confirmed the found duplicates by the traditional Root Mean Square Deviation (RMSD) through optimal alignment.
The DALI took about 30 min on average to find a short list of nearest neighbors of one chain in the whole PDB.
Extrapolating this time to all pairwise comparisons for 707K+ cleaned chains yields 40+ years, slower by orders of magnitude than 6 hours needed for all comparisons of the complete invariants $\bri$ on the same desktop computer.
Our implementation of $\RMSD$ for Fig.~\ref{fig:near-duplicates} has the median time of 534 microseconds per pair of chains (of the same length), about 10 times slower than $L_\infty$ on $\bri$s.
\medskip

The FoldSeek algorithm \cite{van2024fast} is claimed to be 4000 times faster than RMSD by Dali due to optimal alignments of 3-residue subchains instead of full original chains, which takes 3.65 days by the estimates above, still an order of magnitude slower than $L_\infty$ on $\bri$s.
But any similarity distance needs a proof of all metric axioms for trustworthy clustering \cite{rass2024metricizing}. 
\medskip

The ultra-fast speed of all-vs-all comparisons by $\bri$ is explained by the hierarchical nature of this complete invariant.
To find near-duplicates in the PDB, we first compared only average invariants $\brain(S)\in\R^9$.
By Lemma~\ref{lem:brain} the full comparisons by $\bri$ are needed only for a tiny proportion of backbones with the closest vectors $\brain(S)$.
This hierarchical speed-up is unavailable for any distance without underlying invariants.

\section{Discussion of the PDB and data integrity}
\label{sec:discussion}

Using protein structures as an important example, this paper advocates a
justified approach to any real data objects.
The first and often missed step is to define an \emph{equivalence relation} for given data because real objects can be digitally represented in (usually infinitely) many different ways.
\medskip

For instance, a human can be recognized in a huge number of digital photos but science progressed to discover the human genome and other biometric data, which are being included even in passports.
All other objects (protein backbones for example) similarly need complete invariants for unambiguous identification because a distance metric alone is insufficient to understand deeper relations beyond pairwise similarities.
\medskip

There is little sense in distinguishing most objects (including flexible molecules) under rigid motion because translations and rotations preserve their properties in the same environment.
Hence the input of all prediction algorithms should be invariant, ideally a complete continuous invariant. 
\medskip

The Lipschitz bi-continuity is essential because small perturbations of input should not drastically change the output and vice versa.
Weaker versions of Problem~\ref{pro:map} were solved for low-dimensional lattices \cite{bright2023continuous,kurlin2024mathematics}, periodic point sets \cite{edelsbrunner2021density,anosova2021isometry,anosova2026recognition}, sequences \cite{anosova2022density,anosova2023density,kurlin2025complete}, crystals \cite{widdowson2022average,mcmanus2025computing,widdowson2025higher,widdowson2026pointwise}, and finite clouds of unordered points \cite{kurlin2024polynomial,widdowson2023recognizing} in Geometric Data Science \cite{anosova2021introduction,anosova2025geometric,anosova2024importance}. 
\medskip

\textbf{The crucial novelty} in the proposed approach is treating the \emph{rigid class} of any experimental structure (protein backbone) from the PDB as an \emph{objective ground truth} instead of labels or classes assigned manually or by black-box algorithms with many parameters.
Problem~\ref{pro:map} asked for an analytically defined invariant $I$ whose explicit formula will remain the same for any new data without re-training required in machine learning.  
\medskip

While traditional approaches explored finite datasets within infinite spaces in a `horizontal' way, solutions to Problem~\ref{pro:map} and its analogs for other data \cite{kurlin2023strength,kurlin2023simplexwise,anosova2026seeing} provide `vertical' breakthroughs by building geographic-style maps of continuous data spaces as viewed from a satellite \cite{bright2023geographic,widdowson2024continuous,widdowson2025geographic}.
\medskip

Fig.~\ref{fig:PDB707K_BRI_heatmaps_log} and \ref{fig:PDB707K_BRI_heatmaps_by_chain_log} can be zoomed at any spot and mapped by using further invariants.
Such navigation maps with invertible coordinates enable inverse design while any dimensionality reduction to a latent space was proved \cite{landweber2016fiber} to be discontinuous (making close points distant) or collapsing an unbounded region to a point (losing an infinite amount of data).
\medskip

\textbf{The main contributions} are Theorems~\ref{thm:motion}, \ref{thm:continuity}, and \ref{thm:inverse}, which solved Problem~\ref{pro:map} for protein backbones, detected thousands of (near-)duplicates in the PDB and enabled a justified exploration of the protein universe. 
\medskip

The supplementary data (available by request) include the Python code and a table of exact duplicates whose corresponding coordinates coincide in all decimal places and hence might need further refinement. 
The follow-up work \cite{wlodawer2025duplicate} discusses specific duplicates in detail.
Improving the PDB validation is needed to avoid unjustified predictions and claims of `solutions' based on skewed data \cite{mcdonnell2024structure}.
Another analysis \cite{wlodawer2024waterless} raised further concerns by revealing large numbers of waterless structures in the PDB.

\acknowledgment{
This work was supported by the Royal Academy of Engineering Fellowship IF2122/186,  EPSRC New Horizons EP/X018474/1, and Royal Society APEX fellowship APX/R1/231152. 
The authors thank Mariusz Jaskolski, Alex Wlodawer, and Daniel Rigden for their helpful comments on early drafts and all other reviewers for their valuable time.
}

\renewcommand{\thesection}{\Alph{section}}
\setcounter{section}{0}
\section{Appendix: updates on PDB duplicates }
\label{sec:appendix}

\begin{figure*}[h!]
\includegraphics[width=\linewidth]{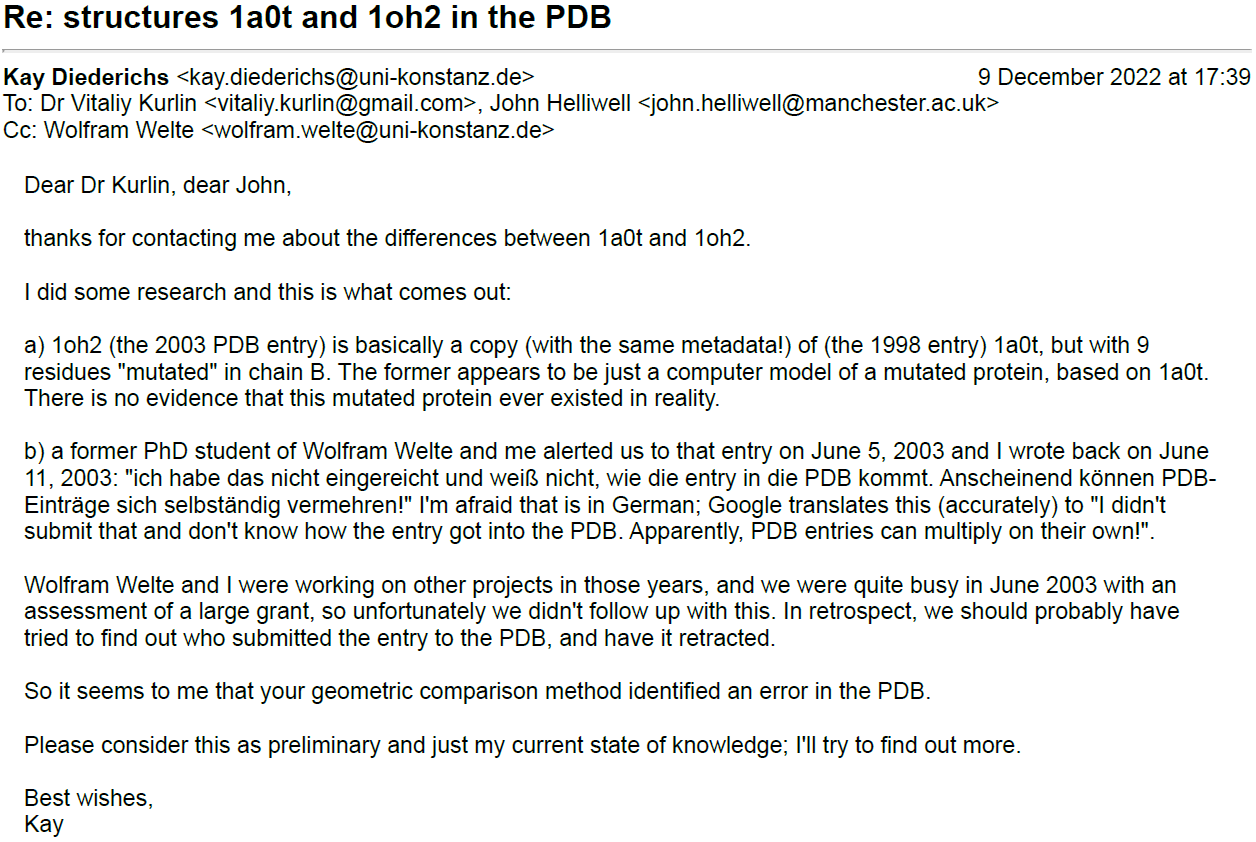}
\caption{Author's confirmation of the duplicates 1a0t and 1oh2.} 
\label{fig:1a0t-vs-1oh2}
\end{figure*}

This appendix discusses several duplicates that were found by the new invariants and later confirmed by their authors, and subsequent updates in the PDB. 
After finding the first duplicates in Table~\ref{tab:duplicates}, we emailed the authors of the underlying publications whose contact details were still possible to find.
The common author of the PDB entries 1a0t and 1oh2, Kay Diederichs, confirmed the error in December 2022 (see Fig.~\ref{fig:1a0t-vs-1oh2}).
\smallskip

\begin{figure*}[h!]
\includegraphics[width=\linewidth]{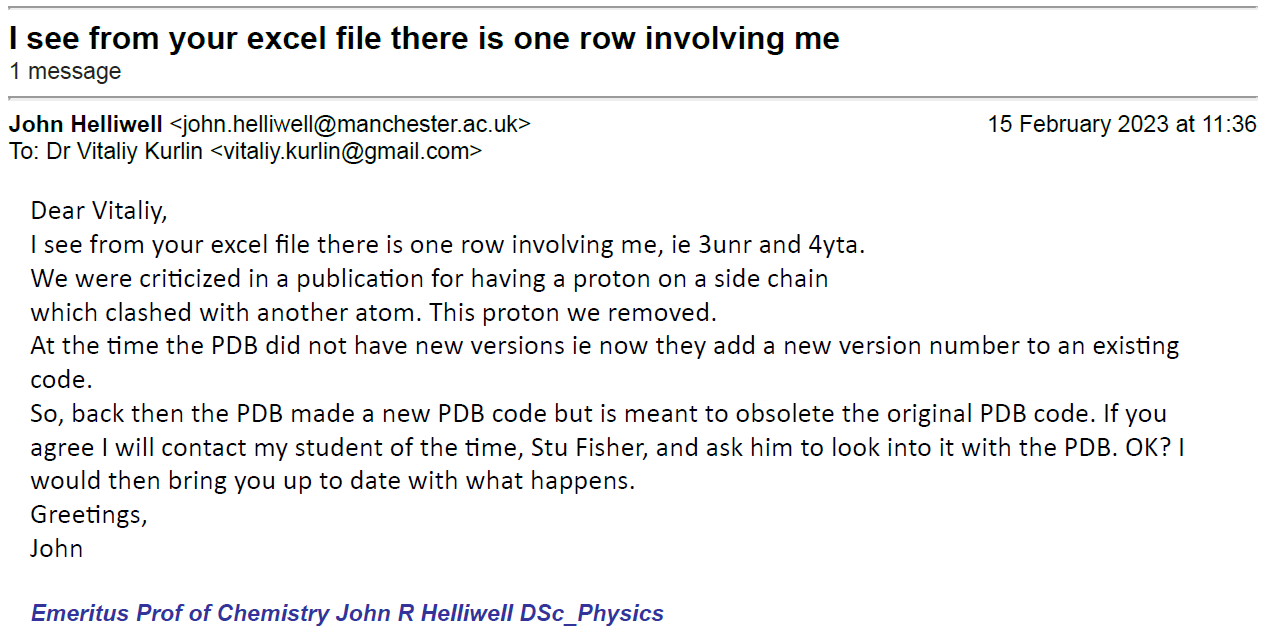}
\caption{Author's confirmation of the duplicates 3unr and 4yta. 
}
\label{fig:3unr-vs-4yta}
\end{figure*}

John Helliwell studied our duplicates including those with the same sequences of amino acids.
After finding his pair of duplicates, he e-mailed us to confirm this error on February 15, 2023 (see Fig.~\ref{fig:3unr-vs-4yta}). 
After meeting with the PDB validation team on February 27, 2023, where John was also present, the webpage https://www.rcsb.org/structure/removed/3UNR was updated without any reference to our work reporting the error: ``Entry 3UNR was removed from the distribution of released PDB entries (status Obsolete) on 2023-03-01. It has been replaced (superseded) by 4YTA''.
\smallskip

The PDB validation team confirmed that PDB entries are updated only by authors' request or by their permission.
After we e-mailed all authors of the first found duplicates in December 2022, five entries from our list were updated in the PDB without acknowledging our work, see Table~\ref{tab:PDBupdates}.
\smallskip

\begin{table}[h!]
  \caption{These five PDB entries had duplicates similar to Table~\ref{tab:duplicates} but were modified after our initial contacts in December 2022.
All original and updated files are still accessible online.}
\label{tab:PDBupdates}
  \centering
  \begin{tabular}{l|ll}
PDB entry & date of modification & reason of modification \\
\hline
4rhv & 2023-01-18 
& Remediation \\
1ruj & 2023-01-18 
& Remediation \\
1gli & 2023-02-08 
& Remediation \\
3hhb & 2023-02-08 
& Remediation  \\
1cov &  
2023-04-19	& Remediation
\end{tabular}
\end{table}

The older versions of all PDB files should be available via the web link \\ ftp://snapshots.rcsb.org/20230102.
Other duplicates in Table~\ref{tab:duplicates} were not previously reported, so their PDB files still show the duplication of geometry with differences in sequences (checked on December 19, 2024).
\medskip

Since all $x,y,z$ coordinates in the PDB are given with three decimal places relative to $1\angstrom$, a distance of less than $0.01\angstrom$ is considered negligible. 


\bibliographystyle{match}       
\bibliography{complete-invariants-proteins}

\end{document}